%% file: testbeam-paper.tex
\documentclass[a4paper,11pt]{article}
\pdfoutput=1 
\usepackage{jinstpub} 
\usepackage[binary-units=true,separate-uncertainty=true]{siunitx}
\usepackage{tikz}
\usepackage{caption}
\usepackage{xspace}

\newcommand{\SIERR}[4]{\ensuremath{\num{#1}\pm\num{#2}\,\text{(stat)}\pm\num{#3}\,\text{(syst)}\,\si{#4}}}

\newcommand{\datura}{\textsc{Datura}\xspace}

\title{Test Beam Performance Measurements for the Phase~I Upgrade of the CMS Pixel Detector}

\abstract{
    A new pixel detector for the CMS experiment was built in order to cope with the instantaneous luminosities anticipated for the Phase~I Upgrade of the LHC.
    The new CMS pixel detector provides four-hit tracking with a reduced material budget as well as new cooling and powering schemes.
    A new front-end readout chip mitigates buffering and bandwidth limitations, and allows operation at low comparator thresholds.
    In this paper, comprehensive test beam studies are presented, which have been conducted to verify the design and to quantify the performance of the new detector assemblies in terms of tracking efficiency and spatial resolution.
    Under optimal conditions, the tracking efficiency is \SI{99.95\pm0.05}{\percent}, while the intrinsic spatial resolutions are \SI{4.80\pm0.25}{\micro\meter} and \SI{7.99\pm0.21}{\micro\meter} along the \SI{100}{\micro\meter} and \SI{150}{\micro\meter} pixel pitch, respectively.
    The findings are compared to a detailed Monte Carlo simulation of the pixel detector and good agreement is found.
}


\input{testbeam-paper-authors.tex}
\emailAdd{simon.spannagel@desy.de}

\begin{document}
\maketitle
\flushbottom

\section{Introduction}

The pixel detector of the Compact Muon Solenoid (CMS) experiment at the CERN LHC~\cite{JINST-cms} is a hybrid silicon pixel tracking detector and is crucial for the reconstruction of particle trajectories as well as for the identification of secondary decay vertices~\cite{CMS-TDR-tracker,CMS-TDR-tracker-add}.
The current CMS pixel detector has been designed to withstand an instantaneous luminosity of $\mathcal{L} =  \SI{1e34}{\centi\meter^{-2}\second^{-1}}$ and performs well under these operating conditions~\cite{tracker-reco}.
However, based on the excellent performance of the LHC, it is anticipated that peak luminosities of twice the original design value are likely to be achieved before 2018 and then probably significantly exceeded during the so-called Phase~I period, which runs from 2018 until 2022.
At higher luminosity and increased hit occupancies the present pixel detector would be subject to severe inefficiencies arising from dead time, limited buffering capabilities, and the limited transmission bandwidth of the front-end electronics.
Thus, a new detector, representing an advancement on the current design, which includes new front-end electronics, has been constructed~\cite{pixel-tdr}.

The redesigned readout chips (ROCs) are a key component of the Phase~I pixel detector upgrade, and their functionality, operational reliability, and performance are of paramount importance.
The characteristics of the chip designed for layers 2 -- 4 of the new pixel detector have been thoroughly tested at the DESY II test beam facility, and the results of these measurements are presented in this paper.
Samples of the chip for layer 1 have not been available yet at the time of the measurements.

This document is structured as follows:
Section~\ref{sec:chip} introduces the new CMS pixel readout chip, while Section~\ref{sec:pixelav} provides information on the simulation of CMS pixel detector modules.
The test beam infrastructure and the beam telescope, as well as the analysis strategy and event selection criteria are described in Section~\ref{sec:testbeam}.
The calculation of statistical uncertainties and the sources of systematic uncertainty relevant for the measurements presented are discussed in Section~\ref{sec:uncertainties}.
Sections~\ref{sec:efficiency},~\ref{sec:intrapixel},~and~\ref{sec:resolution} present the results obtained for the tracking efficiency, intra-pixel charge collection and resolution, and the intrinsic detector resolution, respectively.
Finally, Section~\ref{sec:conclusion} summarizes the findings.
More details about each of the analyses presented in this paper can be found in~\cite{thesis-simon}.

\section{The Sensors and Readout Chips of the CMS Phase~I Pixel Detector}
\label{sec:chip}

\begin{figure}[tbp]
  \centering
  \includegraphics[width=.5\textwidth]{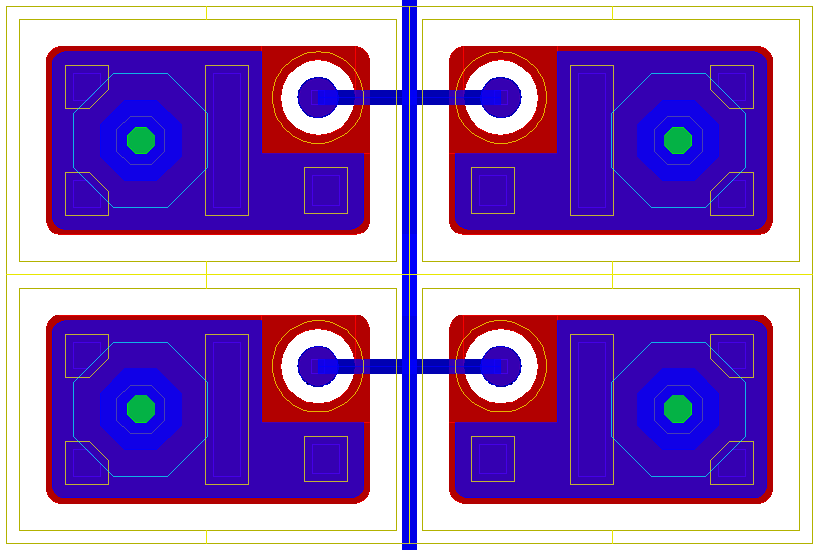}
  \caption[Pixel implant layout for the CMS pixel sensors]{Pixel implant layout for the CMS pixel sensors. The pixels of the sensor design used in the test beam measurements are separated by moderated \emph{p}-spray isolation~\cite{moderated-pspray} with a width of \SI{30}{\micro\meter}. The punch-through grid runs vertically and connects the bias dots in the individual pixels~\cite{tilman-private}.}
  \label{fig:pixelcell}
\end{figure}

The silicon sensors of the new CMS pixel detector implement an \emph{n$^+$}-in-\emph{n} design consisting of \emph{n}-doped silicon substrate with a nominal thickness of \SI{285}{\micro\meter} and highly doped \emph{n$^+$} pixel implants for the electrode segmentation.
The back side is uniformly \emph{p}-doped below the active sensor area to form the backplane electrode.
The isolation of the pixel implants is ensured by additional moderated \emph{p}-spray~\cite{moderated-pspray} or \emph{p}-stop structures surrounding the pixel cells.
For the measurements presented in this paper, sensors with \emph{p}-spray implants have been used.
The punch-through biasing grid visible in Figure~\ref{fig:pixelcell} allows biasing of the sensor without connected readout electronics by providing the ground potential and prevents single pixels from floating in case of missing bump bonds.
The pixel pitch of $\SI{100}{\um} \times \SI{150}{\um}$ is designed for optimal charge sharing between two pixel cells in the presence of the \SI{3.8}{\tesla} magnetic field of the CMS solenoid~\cite{sensors}.

The ROC constitutes the front-end electronics responsible for the detection and transmission of particle hits, and is fabricated using a \SI{250}{\nm} CMOS process with radiation tolerant design.
The chip has a total size of $\SI{7.9 x 10.2}{\mm}$.
The active area accommodates 4160 pixels, which are organized into 26 double columns of $80 \times 2$ pixels each.
Every pixel cell is equipped with all the components necessary for the amplification and discrimination of the signals from the silicon sensor.
The ROC is supplied with a \SI{40}{\MHz} clock enabling beam-synchronous data acquisition at the LHC.
Two independent power supplies for the analog and digital parts of the circuitry minimize the influence of digital signal processing on the analog performance.
A set of digital-to-analog converters are used for the configuration of the chip; communication is implemented via a custom-designed \SI{40}{\MHz} Inter-Integrated Circuit (I$^2$C) interface.

The charge signals collected from the sensor implants are sampled by the on-chip analog-to-digital converter (ADC).
The ADC provides \SI{8}{\bit}s of pulse height information for every pixel hit, which can be used for cluster center interpolation.
An in-time threshold of about \SI{1.7}{ke} is achievable.
The data recorded are transmitted outside the detector via a \SI{160}{\MHz} digital link.
A more detailed description of the new front-end electronics can be found in~\cite{Kästli201388}, while a comparison of the old and the new ROC is provided in~\cite{upgrade-plans}.

The assemblies used for the measurements presented in this paper consist of one ROC and a sensor with a physical size of about $\SI{10 x 10}{\mm}$, while the modules produced for the CMS Phase~I Pixel Detector feature 16 ROCs on one sensor.
The ROCs were bump-bonded to the sensors at DESY using tin-silver alloy solder balls~\cite{panic2014}.

\section{Simulation of CMS Pixel Detector Assemblies}
\label{sec:pixelav}

The passage of particles through the CMS pixel assemblies is simulated using the Pixelav package~\cite{Swartz200388,Swartz:687440}.
The simulation is compared with the test beam data, facilitating understanding of the charge deposition and collection processes in the silicon sensor.

The charge deposition in the sensor is modeled using the elastic pion-electron cross section, which is applied without further modification for data collected using the positron beam at the DESY II test beam facility.
The probability and range of the delta rays as well as the stopping power are taken from the ESTAR database~\cite{estar}.
The transport of the deposited charge carriers is modeled by numerically integrating the relevant equations of motion using a fifth-order Runge-Kutta method~\cite{Swartz:687440}.
An electrostatic simulation of the sensor is performed using Synopsys TCAD~\cite{synopsis-tcad} and provides the electric field required for the time evolution.
In addition, diffusion processes are accounted for by adding a randomized offset to each coordinate of the charge carrier at each integration step.

Simplified models of the readout electronics are used in order to describe the influence of the charge threshold and electronic noise.
The models follow the basic building blocks of the amplifier scheme in the pixel cells and the ADC of the chip, and allow us to describe the features
observed in test beam data well.
The model parameters are tuned to best describe the test beam data.
First, a thermal noise contribution, described by a Gaussian distribution with an RMS of 180 electrons, is added to the charge collected at the sensor implants.
Second, the charge threshold is smeared by a normal distribution with a width of 100 electrons and the signal response of the shaper circuit is modeled using the cumulative Weibull function~\cite{weibull}.

In order to correctly describe the width of the Landau distribution of the cluster charge obtained from test beam data, an additional multiplicative smearing of \SI{6}{\percent} is applied to the preamplifier gain.
Furthermore, pixel charges exceeding the simulated comparator threshold are smeared using a Gaussian distribution with a width of 500 electrons, in order to take into account the limited precision of the pulse height sampling in the \SI{8}{\bit} on-chip ADC.
Finally, the finite track resolution of the beam telescope described in the next section is mimicked by smearing the particle incidence position using normal distributions, each with a width corresponding to the expected track resolution of about \SI{5}{\um}.

\section{Test Beam Setup and Beam Telescope}
\label{sec:testbeam}

The data presented in this paper were recorded at the DESY test beam facility. 
The DESY II synchrotron circulates one electron bunch at a frequency of about \SI{1}{\MHz} with a maximum energy of about \SI{6.3}{\GeV}.
The test beam is generated via a twofold conversion~\cite{eudet-memo-2007-11} and a spectrometer dipole magnet enables momentum selection in the range 1 -- \SI{6}{\GeV}.
The beam has a divergence of about \SI{0.5}{mrad} and an energy spread of approximately \SI{5}{\percent}.
Depending on the beamline and the selected beam momentum, particle rates of a few 10 kHz can be achieved.

The reference tracks for this analysis are provided by the \datura beam telescope~\cite{datura-paper}, a six-plane pixel detector telescope featuring MAPS-type MIMOSA26 sensors with a pixel pitch of $\SI{18.4}{\micro \meter} \times \SI{18.4}{\micro \meter}$~\cite{HuGuo2010480}.
The sensors are thinned down to a thickness of \SI{50}{\micro \meter} to reduce the material budget; the charge carriers are collected from a \SI{20}{\micro\meter} epitaxial layer by diffusion.
The total material budget of the telescope amounts to only \SI{300}{\micro\meter} of silicon sensor material, and \SI{300}{\micro\meter} of protective Kapton foil.
The MIMOSA26 sensors were operated at threshold level 6, i.e., requiring the signal pulse height to be at least six times the noise RMS.
The average noise occupancy for this threshold at room temperature is about \num{6e-5} per pixel~\cite{mimosa26}.
The corresponding intrinsic telescope plane resolution has been measured to be $\sigma_{\mathrm{tel}} = \SI{3.24\pm0.09}{\micro\meter}$~\cite{datura-paper}.
The MIMOSA26 sensors operate in a rolling shutter readout mode with an integration time of about \SI{120}{\micro\second}, which causes track pile-up.

\begin{figure}[tbp]
  \input{telescope}
  \caption[Geometry of the test beam setup]{Geometry of the test beam setup featuring a beam telescope with six tracking planes, scintillators, and the device under test (DUT) between the upstream and downstream arms of the telescope. An additional timing reference detector (REF), similar in nature to the DUT, is placed downstream of the telescope.}
  \label{fig:telescope}
\end{figure}
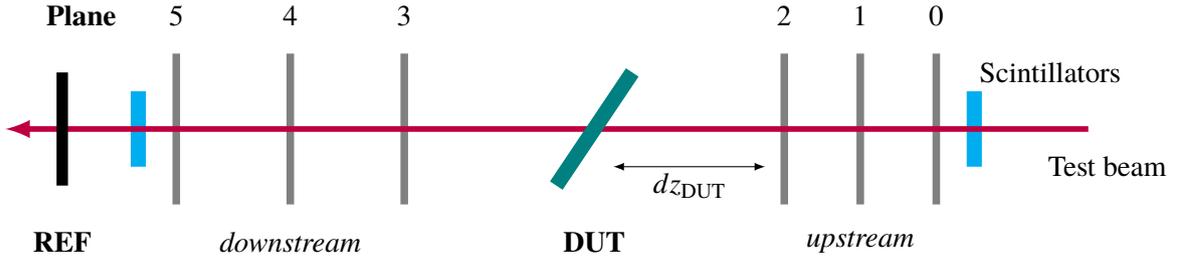

The apparatus used in these test beam measurements consists of the \datura beam telescope; scintillator triggers; and two CMS pixel assemblies, one as the device under test (DUT) and the other one as timing reference (REF), as depicted in Figure~\ref{fig:telescope}.
Four PMT assemblies with plastic scintillators and light guides are used to generate trigger signals when beam particles traverse the setup.
The scintillators are mounted pairwise before the first and after the last telescope plane, respectively, forming a trigger window of approximately $\SI{10}{\milli\meter} \times \SI{10}{\milli\meter}$, which matches the size of the CMS pixel assemblies.

The ROC and sensor assemblies of the DUTs and the REF are glued to small PCB carrier boards with edge connectors in order to  allow testing and fast sample change in the beam line.
The carrier board is attached to a PCB mounted on a copper plate and connected to the readout electronics.
The copper plate features a cut-out around the position of the DUT in order to reduce the material present in the beam.
A cooling loop inside the plate allows coolant from an ethanol-based chiller to circulate in order to stabilize the sample at room temperature or to operate at temperatures down to \SI{-20}{\celsius}.
The assembly can be encased by a polyethylene cover that permits flushing with dry air to avoid condensation.
For high-resolution measurements of unirradiated assemblies, the DUT is operated without the cover to further reduce the material budget and to minimize the distance to the upstream telescope plane.
In this configuration, the DUT is stabilized at a temperature of \SI{17}{\celsius}.

Owing to the rolling shutter readout of the MIMOSA26 sensors, the mean track multiplicity per event in the \datura telescope is up to 10 depending on the beam energy and intensity.
Tagging of the correct track arriving within the \SI{25}{\nano\second} trigger window of the CMS pixel devices is accomplished using the timing reference detector, which is mounted downstream of the telescope, as indicated in Figure~\ref{fig:telescope}.
In order to emulate beam-synchronous operation of the detectors, triggers are only accepted within the first \SI{8}{\ns} of the \SI{40}{\MHz} clock supplied to the CMS pixel assemblies.

The data acquisition is performed using the EUDAQ~\cite{EUDET-2010-01} software framework and the pxarCore library for the CMS pixel detectors~\cite{pxar-manual}.

\subsection{Offline Reconstruction}
\label{sec:reco}

The EUTelescope software~\cite{EUDET-2010-12} is employed for particle reconstruction.
First, the reference tracks from the \datura telescope are reconstructed and the telescope is aligned.
Then, the tracks are matched to the measurements from the DUT and REF detectors and their alignment is determined.

The binary pixel hit information provided by the MIMOSA26 sensors is cleared of noisy pixels by applying a threshold on the firing frequency for individual pixels.
Adjacent pixel hits are combined into clusters, and all clusters containing a noisy pixel are removed from the subsequent analysis.
The cluster position is transformed from the local sensor frame of reference to the global beam telescope coordinate system for alignment.

The telescope planes are aligned using a two-step method.
In the first step, a pre-alignment based on correlation histograms for the different telescope planes is performed.
The hits in $x$ and $y$ of all neighboring telescope plane combinations are correlated and the resulting residual distributions are shifted to minimize their means.
The second step requires full telescope tracks to be reconstructed, which are then passed to the \textsc{Millepede-II} algorithm~\cite{Blobel2011, millepede}.
\textsc{Millepede-II} implements a global $\chi^2$ minimization approach and returns the alignment parameters and uncertainties for each of the telescope planes.
The first and second telescope planes are fixed in their initial positions to define the global coordinate system.
These two planes were used in order to exclude any influence from additional scattering in the material of the DUT, which is located between the two telescope arms.

\begin{figure}[tbp]
  \centering
  \includegraphics[width=.5\textwidth]{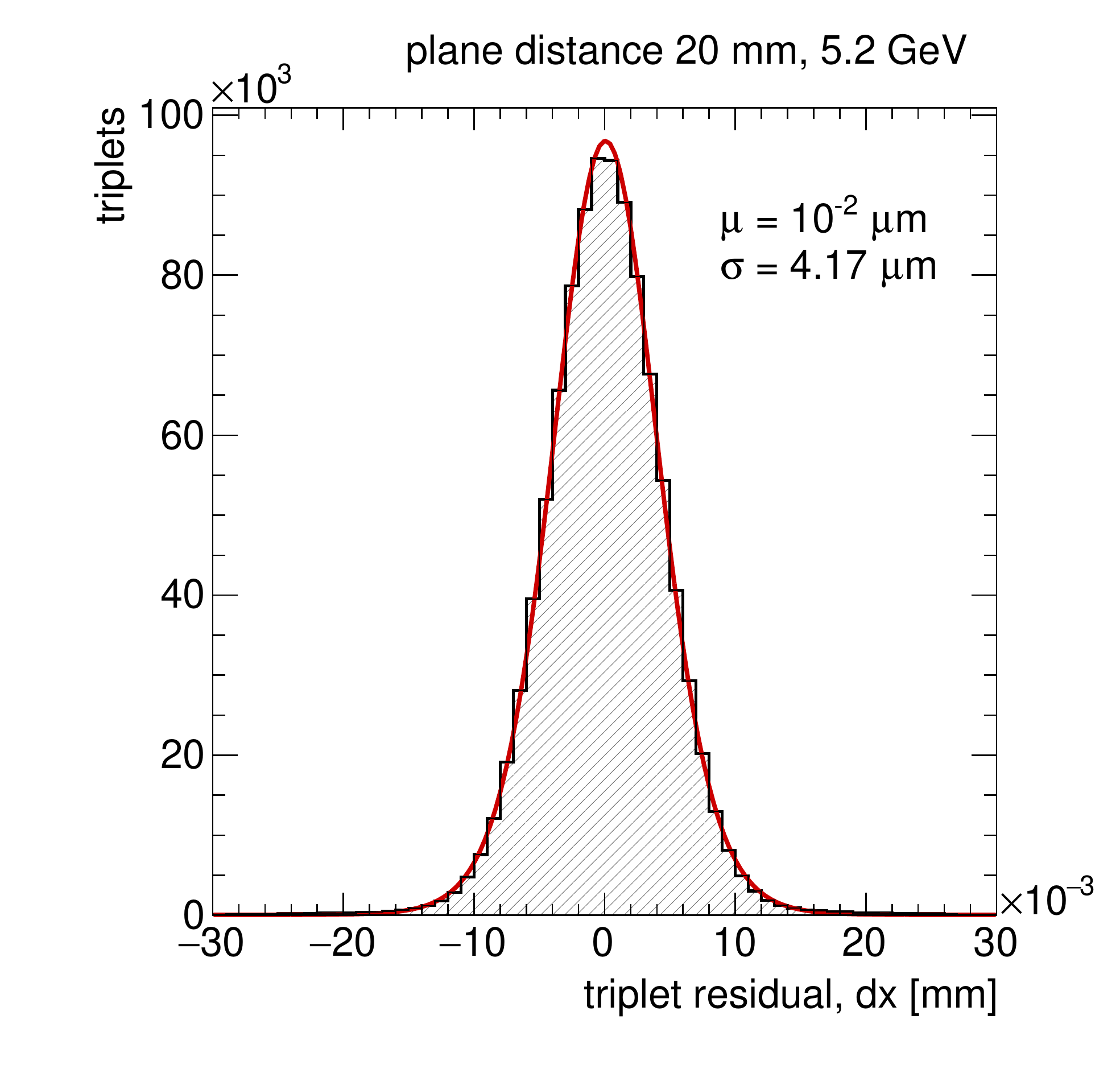}%
  \includegraphics[width=.5\textwidth]{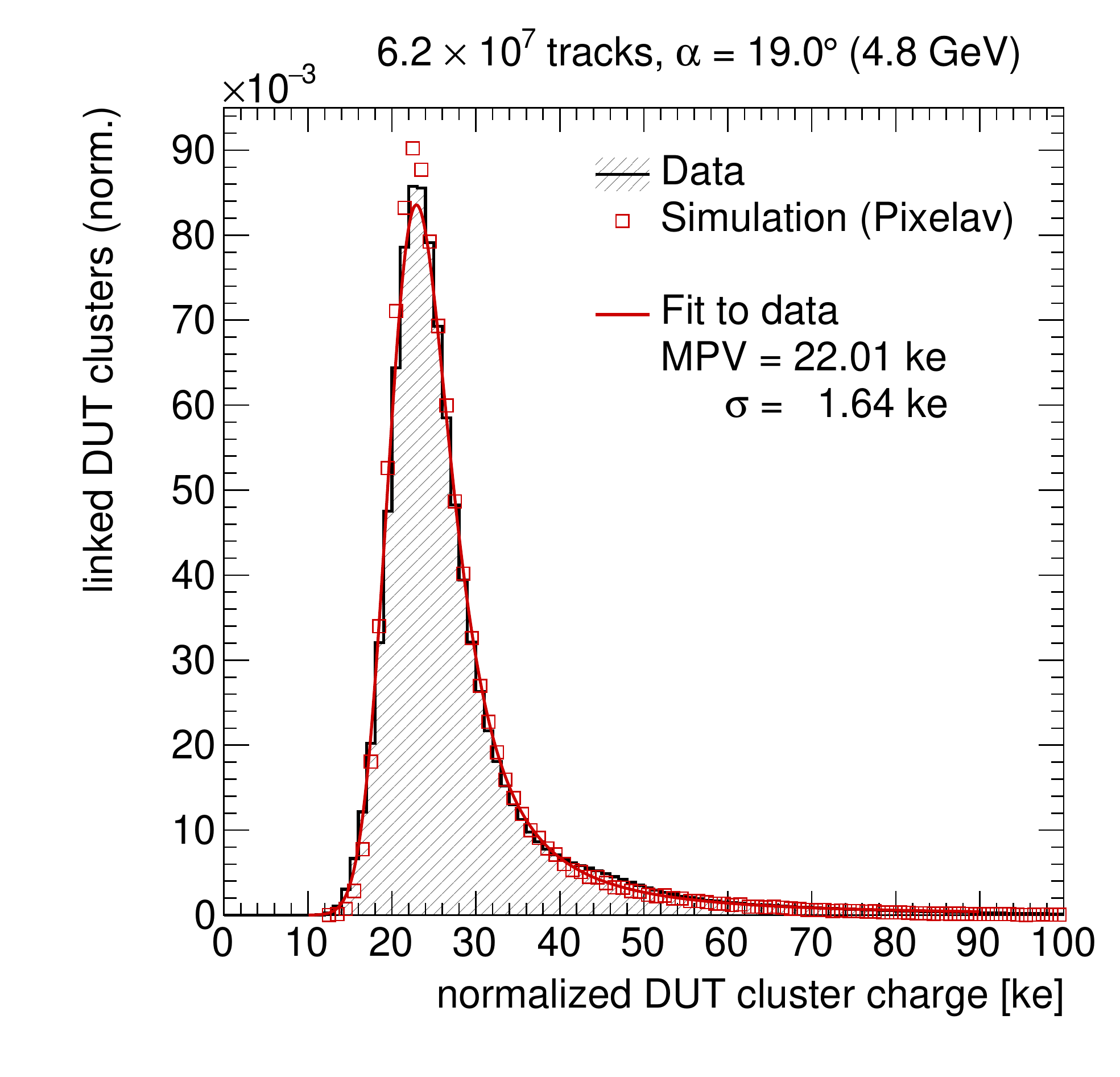}%
  \caption[Normalized cluster charge]{Unbiased residual distribution of the upstream arm triplet at telescope plane 1~(left). Cluster charge distribution normalized to vertical track incidence~(right). Only DUT clusters matched to telescope tracks in the fiducial volume (cf.\ Section~\ref{sec:selection}) are shown.}
  \label{fig:cmsq0f}
\end{figure}

Track finding is performed using the triplet method.
A track in the \datura telescope is required to have exactly one associated hit in each of the six planes.
The initial track candidate is built from track triplets formed independently in the upstream and downstream arms of the beam telescope.
These track candidates with their initial residuals in each plane are then fitted using the General Broken Lines (GBL) method~\cite{Blobel2006,Kleinwort2012107}.
Triplets in the upstream telescope arm are formed by first considering hits in telescope planes~0~and~2 only.
For all possible hit combinations, the straight line connecting these two measurements is then interpolated to the position of telescope plane~1.
If a matching hit on plane~1 can be found within \SI{200}{\micro\meter} of the interpolated impact point, the hit combination is accepted as a track triplet.
The same procedure is repeated for telescope planes~3~and~5, which require a matching hit in plane~4 within the same distance of the triplet candidate.
The two sets of triplets are then extrapolated to the nominal $z$-position of the DUT in the center of the telescope and all combinations of upstream and downstream triplets are compared for possible intersection.
If an intersection point can be found within \SI{300}{\micro\meter}, the triplet pair is accepted as a track candidate.
GBL trajectories are built from the six measurements of the track candidate, while the upstream triplet is used as a track seed.
In addition to the actual measurements, scattering material in the telescope planes as well as the DUT are introduced to account for multiple Coulomb scattering.
For some measurements, reference tracks formed only from the upstream triplets are used in order to exclude scattering in the material of the DUT installation.
The unbiased triplet residual at the center plane of the upstream telescope arm after alignment is shown in Figure~\ref{fig:cmsq0f}~(left).
A fit of the residual distribution with a generalized error function yields a width of \SI{4.17}{\um}, which translates to an intrinsic plane resolution of about \SI{3.4}{\um} after the propagation of uncertainties.
Further information on the alignment procedure and the precision of the beam telescope, as well as detailed measurements of the intrinsic resolution can be found in~\cite{datura-paper}.

The data from the CMS pixel assemblies are reconstructed and calibrated as follows.
The \SI{8}{\bit} charge information provided by the on-chip ADC described in Section~\ref{sec:chip} and~\cite{Kästli201388} is converted into units of the internal calibration pulse using the inverse of the cumulative Weibull function employed for the calibration of the chip response.
The absolute charge calibration relating the internal calibration pulse to the number of electrons deposited in the sensor is obtained from simulation.
A convolved Landau-Gauss distribution is fitted to the cluster charge spectrum, and the conversion factor is chosen such that the most probable value (MPV) of the charge distribution matches the value obtained from fitting the same distribution to the simulation described in Section~\ref{sec:pixelav}.
The calibrated cluster charge distribution as well as the fit results are shown in Figure~\ref{fig:cmsq0f}~(right).
Here, the cluster charge is normalized to vertical track incidence using the relation $Q_0 = Q\cos{\alpha}\cos{\omega}$, where $Q$ is the total cluster charge while $\alpha$ and $\omega$ denote the Euler angles in row and column direction, respectively.
The small deviation between data and simulation visible around \SI{45}{ke} arises from small deviations in the charge calibration at the saturation point of the cumulative Weibull function.

A sparse clustering algorithm is used to form clusters from the DUT and REF pixel hits.
Starting from a seed pixel, the cluster is extended by calculating the distance of all remaining pixels to the pixels already assigned to the cluster.
Pixels found within a distance of one pixel are added to the cluster and removed from the list of remaining, unclustered pixel hits.
If no more adjacent pixels can be found, the cluster is marked as complete, and a new seed pixel is selected.
The position of the cluster center is calculated with the Center-of-Gravity algorithm, which makes use of the available pixel charge information~\cite{cog}.
The reference tracks (reconstructed in the beam telescope as described above) are transformed into the DUT frame of reference using a passive coordinate transformation.
This facilitates the comparison with sensor features such as track impact position relative to the pixel pitch.
Pixel hits are reconstructed on the middle plane of the sensor volume.

The alignment of the DUT and REF detectors is performed in their respective local reference frames using the same iterative alignment procedure as employed for the beam telescope, with a pre-alignment being performed before the \textsc{Millepede-II} alignment step.
The DUT is aligned in all three spatial directions, $a_x$, $a_y$ and $a_z$, and the three Euler angles, $\alpha$, $\omega$ and $\phi$, while the REF detector at vertical incidence only requires alignment in the spatial directions and the rotation $\phi$ around the beam axis.
The alignment procedure described above determines shifts with a precision in the sub-micron range and the rotation angles of the DUT with a precision better than~\SI{0.1}{\degree} for all angles studied.

\subsection{Event Selection}
\label{sec:selection}

A set of selection criteria are imposed on the reconstructed test beam data in order to maximize the sensitivity of the analysis to the features of the assembly under test.

The sensors of the CMS pixel detector contain enlarged edge pixels in order to fit the dimensions of the ROC and to add an additional placing margin between neighboring chips on a module without creating insensitive areas.
The sensors of the CMS pixel assemblies used for the test beam measurements feature the same enlarged pixels at the edges and corners of the chip.
Since the electric field and charge collection behavior are different in these pixel cells, clusters extending into these regions are removed from the analysis, and the remaining sensor volume is referred to as the fiducial volume.

Acceptance criteria imposed on the telescope reference tracks help to significantly improve the resolution.
The telescope tracks are required to have a clearance of more than \SI{300}{\micro\meter} in $x$ and $y$ at the position of the DUT without another telescope track present.
This isolation requirement reduces matching ambiguities.
Tracks from particles that have undergone heavy scattering or do not originate from the beam are filtered out based on the slope of the reference tracks.
The slope in $\theta_x$ and $\theta_y$ of the telescope upstream triplet are considered and tracks with a slope $\theta > \SI{2}{\milli\radian}$ are rejected.

\begin{figure}[tbp]
  \centering
  \includegraphics[width=.5\textwidth]{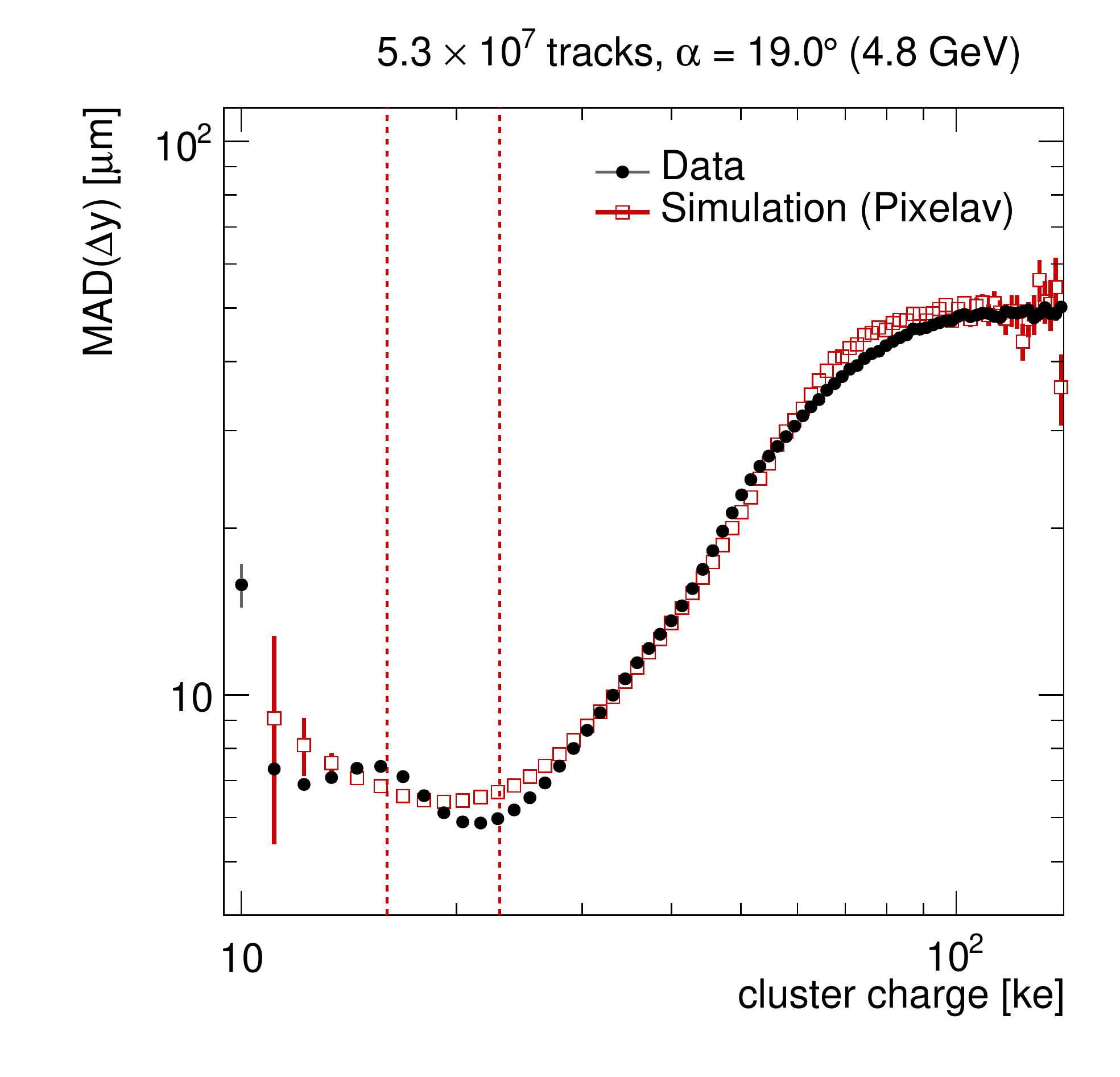}
  \caption[Residual width as a function of the cluster charge]{Residual width as a function of the cluster charge. The influence of delta rays on the residual distribution can be observed at large cluster charges. Dashed lines mark the $Q_0$ selection requirement on the cluster charge normalized to the vertical track incidence angle.}
  \label{fig:cmsrmsvsq}
\end{figure}

Only DUT clusters that can be matched to a telescope reference track within certain limits are accepted.
Delta rays produced in the sensor severely affect the position resolution as the delta electron created at the point of incidence with the sensor can travel significant distances within the sensor volume, producing large clusters.
Figure~\ref{fig:cmsrmsvsq} shows the residual width, measured as the mean absolute deviation (MAD), as a function of the cluster charge.
A deterioration towards large cluster charges is visible, while the best resolution is achieved around the MPV of the cluster charge distribution.
The background stemming from delta rays can therefore be efficiently rejected by imposing a quality requirement on the cluster charge around the Landau MPV value.
The deposited charge (and thus the Landau MPV) depend on the track incidence angle, hence the selection criterion is applied to the cluster charge normalized to the vertical track incidence, $Q_0$.
All clusters with $\SI{16}{\kilo e} < Q_0 < \SI{23}{\kilo e}$ are selected as indicated by the dashed lines in Figure~\ref{fig:cmsrmsvsq}.

\section{Statistical and Systematic Uncertainties}
\label{sec:uncertainties}

The statistical uncertainty on the resolution measurements is calculated using pseudo-experiments.
The number of entries in each bin of the residual distribution under consideration is smeared with a Poisson distribution
whose mean value is set to that of the original bin content.
The residual width obtained from the smeared histogram is stored, and the pseudo-experiment repeated \num{10000} times.
The statistical uncertainty on the measured residual width is then taken as the width of the resulting distribution and is propagated to the intrinsic resolution.

Systematic uncertainties for the measurements presented are mainly related to the environmental conditions during the test beam such as temperature or beam parameters, and the characteristics of the \datura beam telescope used for the reference tracks.
Most systematic uncertainties enter the measurement via the calculated telescope track resolution, which is subtracted quadratically from the measured residual width in order to obtain the intrinsic resolution of the detector.
The track resolution to be subtracted is calculated using the GBL track resolution calculator~\cite{gbltool}.
The input parameters to the track model are the beam energy; the distance, $dz_{\mathrm{DUT}}$, between the telescope planes and the DUT detector; the intrinsic resolution and material budget of the telescope planes; and the material budget of the DUT installation itself.
The uncertainty on the track resolution arising from both the beam energy and $dz_{\mathrm{DUT}}$ can be estimated by varying the corresponding parameters within their uncertainties in the calculation.
Considering tracks from the upstream triplet only (so as to maximize the uncertainty) and a measured DUT distance of around $dz_{\mathrm{DUT}} \approx \SI{45}{\milli\meter}$, the uncertainty stemming from the beam energy is below \SI{\pm0.10}{\micro\meter} for measurements at \SI{5.2}{\GeV}.
At a beam energy of \SI{5.6}{\GeV}, which was used for some of the measurements, the uncertainty is expected to be even lower.
With a rather conservative estimate of the uncertainty on the distance measurement between the telescope upstream arm and the DUT of $\Delta dz_{\mathrm{DUT}} \approx \SI{2}{\milli\meter}$, the effect on the extrapolated track resolution is estimated to be below $\sigma = \SI{\pm0.10}{\micro\meter}$.

The systematic uncertainty arising from the intrinsic resolution on the track extrapolation can be estimated by repeating the GBL calculations using different values for the intrinsic plane resolution.
A variation of the resolution within its uncertainty yields a track resolution uncertainty of about \SI{\pm0.09}{\micro\meter} for tracking using the three upstream planes only, and about \SI{\pm0.07}{\micro\meter} when all six telescope planes are used.
Variations in the detector alignment are of no concern since the data acquisition was limited to short runs and a re-alignment of the DUT as well as the REF detectors was performed for every run.
The total systematic uncertainty on the predicted track resolution is calculated as the quadratic sum of the individual sources and amounts to $\Delta \sigma_{\mathrm{track}} = \SI{\pm0.17}{\micro\meter}$.

The material budget of the DUT is relevant for the telescope track resolution calculated using all six planes.
However, this is only necessary for efficiency measurements where track tagging with the REF detector is required, and resolution measurements are performed using three-plane telescope tracks only.
For efficiency measurements, the track resolution is of no concern as very loose DUT matching criteria are applied, only requiring a telescope track within \SI{1}{\milli\meter}.

Temperature changes of the ROC affect the response behavior of the ADC by \SI{150}{electrons \per \celsius} and thus the charge calibration.
Since environmental temperature changes are rather slow, no changes are to be expected between runs, but may appear for measurements taken on different days.
The offset in the Landau MPV has been corrected by applying different calibration factors (which convert from internal calibration pulse units to electrons) to these sets of runs using the MPV of the charge distribution simulated by Pixelav as reference as described in Section~\ref{sec:reco}.
In order to account for any remaining dependencies and the uncertainty arising from the simulation, the rejection criteria for the cluster charge are shifted by \SI{\pm1}{\kilo e} and the resolution measurement is repeated.
The uncertainty on the measured resolution is found to be $\Delta \sigma_{\mathrm{meas}} = \SI{\pm0.12}{\micro\meter}$, averaging over all measurements at all track incidence angles.

Efficiency measurements are affected by the absolute particle rate and timing instabilities associated with the phase of trigger arrival relative to the detector clock.
The particle rate influences the internal buffering mechanisms of the ROC and might lead to inefficiencies whenever buffer overflows occur.
However, at the DESY II test beam rates of only a few \si{\kHz \per \square \centi\meter} can be achieved.
With the chip being designed for particle rates of \SI{150}{\MHz \per \square \centi\meter} and above, these inefficiencies can safely be neglected.
The uncertainties arising from trigger timing instabilities are difficult to quantify and are assumed be of the order of the measured inefficiencies of around 0.5\textperthousand.

Modeling uncertainties associated with the simulation are evaluated by altering the simulated sensor thickness by \SI{\pm8}{\um} and the applied charge threshold by \SI{\pm0.5}{ke}.
The maximum difference to the nominal simulation is taken as the uncertainty, and the individual contributions are summed in quadrature.
It should be noted that the simulations only enter the measurements through the absolute charge calibration.

\section{Tracking Efficiency}
\label{sec:efficiency}

The tracking efficiency of the CMS pixel ROC has been measured as a function of operation time, telescope track multiplicity and track position on the chip at a comparator threshold of \SI{1.7}{ke}, and is calculated using six-plane telescope tracks that have been tagged by the REF detector as being in the correct \SI{25}{\nano\second} trigger window.
The DUT and REF detectors are operated synchronously using the same externally generated \SI{40}{\MHz} clock, and triggers are only accepted within a window of about \SI{8}{\nano\second} after the rising clock edge to ensure that the trigger timing is correct, as described in Section~\ref{sec:testbeam}.
The tracking efficiency is then defined as the fraction of tracks that pass the isolation criteria, and which have an associated cluster in both the REF and DUT detectors, over the total number of isolated tracks with matching clusters in the REF plane.
\begin{figure}
  \centering
  \includegraphics[width=.5\textwidth]{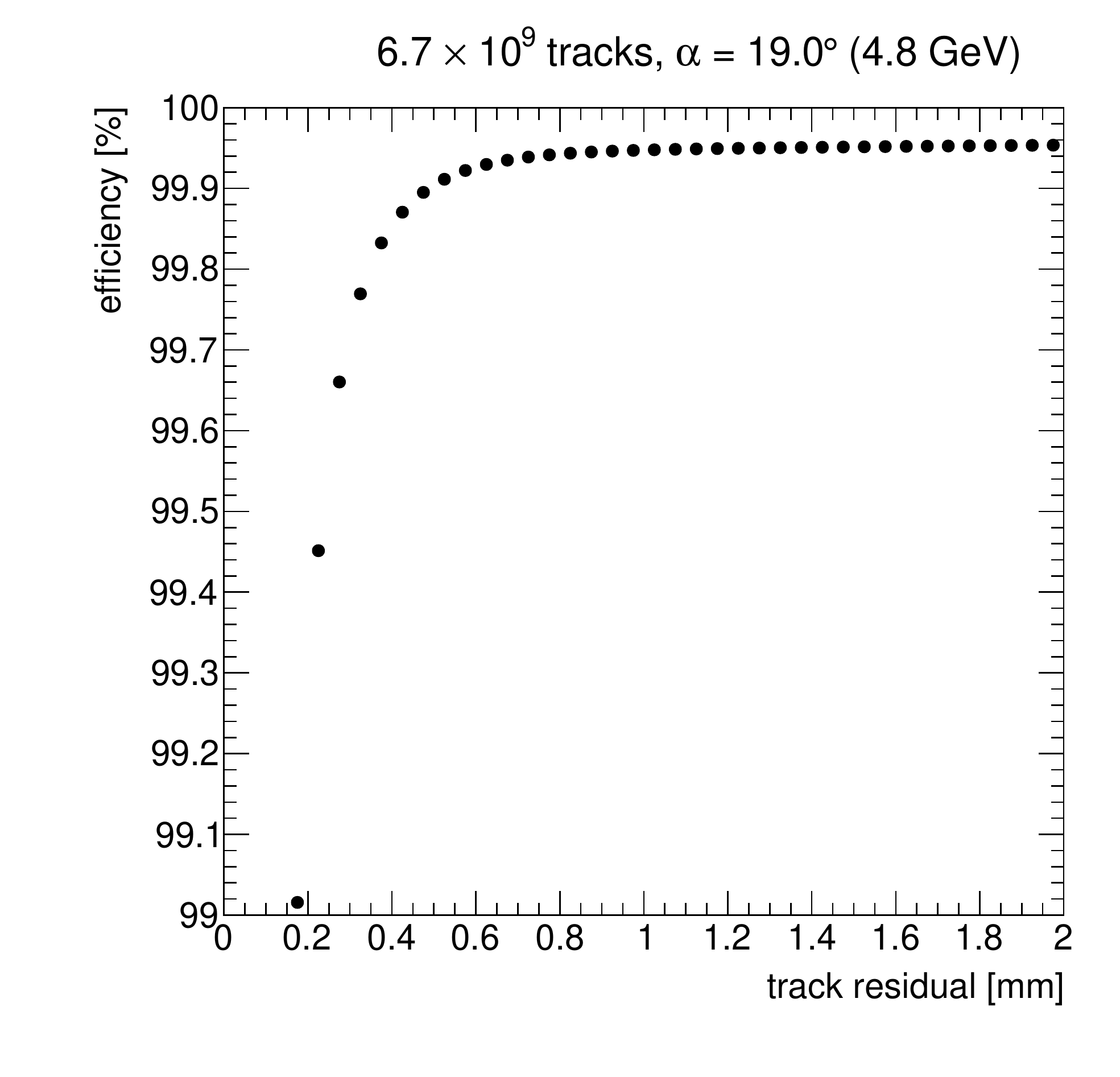}%
  \includegraphics[width=.5\textwidth]{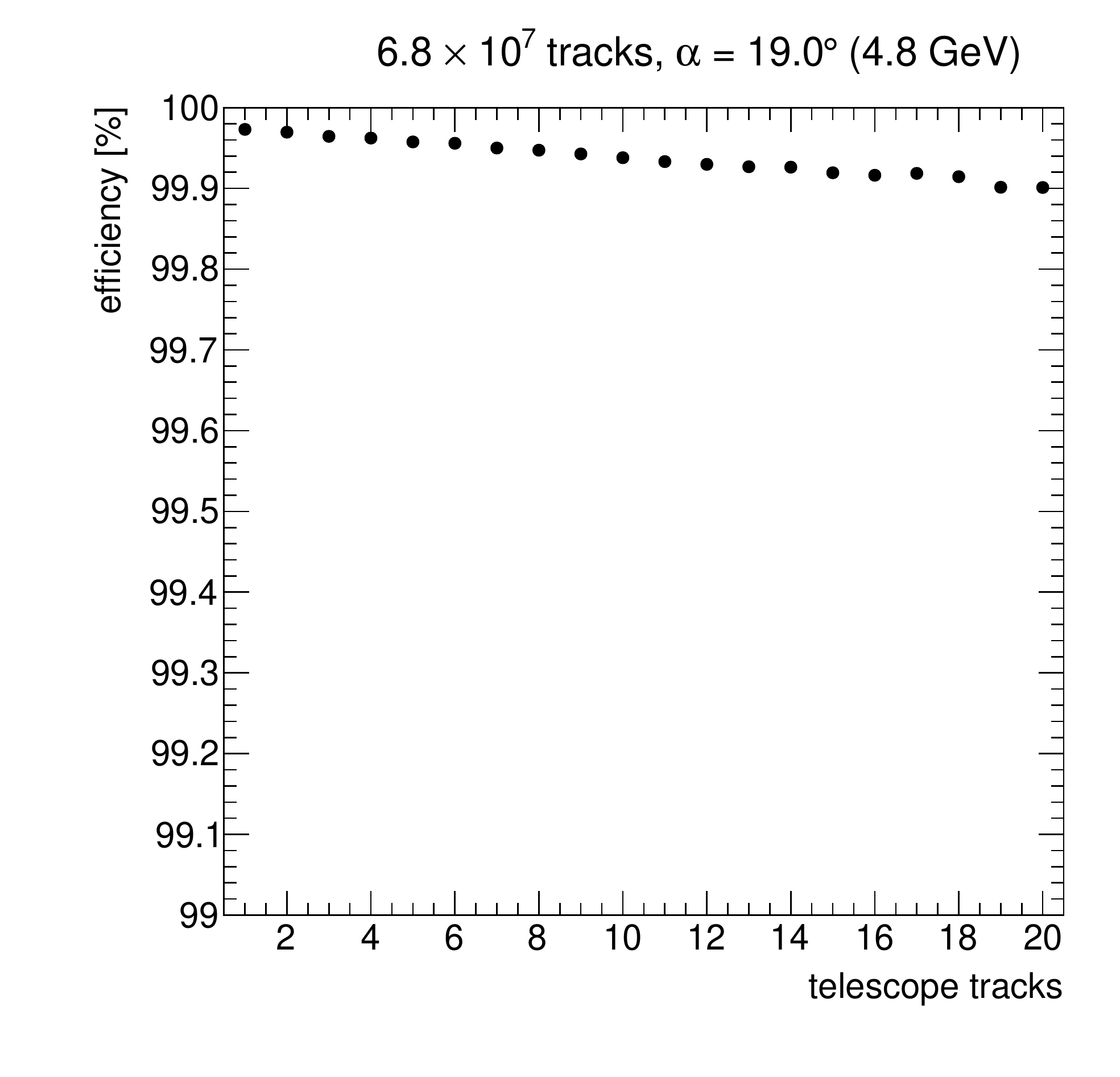}
  \caption[Tracking efficiency as a function of the acceptance window and the telescope pile-up]{DUT tracking efficiency as a function of the size of the track acceptance window around the cluster center~(left) and as a function of the telescope reference track pile-up~(right).}
  \label{fig:effxy}
\end{figure}
Figure~\ref{fig:effxy}~(left) shows the efficiency as a function of the dimension of the square acceptance window. Tracks with a residual within the acceptance window are matched with the measured DUT cluster, while tracks with larger residuals are rejected.
The efficiency saturates at track residuals of about \SI{1}{\mm}. These residuals stem from large clusters produced by delta rays, which significantly shift the reconstructed cluster center.

The measured efficiency exhibits a mild dependency on the telescope track multiplicity due to matching ambiguities, as demonstrated in Figure~\ref{fig:effxy}~(right).
It shows the tracking efficiency achieved with the DUT versus the number of particle tracks present in the telescope event.
With 20~telescope tracks present, an efficiency of \SI{99.90}{\percent} can be reached, while restricting the selection to events with only a single telescope track yields an efficiency of about \SI{99.97}{\percent}.
No dependency on the running time has been observed.
The averaged efficiency over the full fiducial volume of the sensor, including all events selected following the procedure described in Section~\ref{sec:selection}, and with an acceptance of track residuals smaller than \SI{1}{\mm} is evaluated to be
\begin{equation*}
  \epsilon = \SIERR{99.95}{0.01}{0.05}{\percent}\text{,}
\end{equation*}
with the systematic uncertainties estimated as described in Section~\ref{sec:uncertainties}.
A cross-check measurement has been performed using the DUT detector as the timing reference while measuring the tracking efficiency of the REF plane.
Results similar to those determined using the DUT measurement were achieved.

\section{Intra-Pixel Resolution and Efficiency}
\label{sec:intrapixel}

Quantities such as spatial resolution, charge collection or tracking efficiency can be measured as a function of the intra-pixel track position using the high-resolution reference tracks provided by the \datura beam telescope.
All plots presented in this section depict a $2\times2$ pixel array matching the drawing presented in Figure~\ref{fig:pixelcell}.
The measurements with matched tracks from the full fiducial volume are projected into these pixel cells in order to increase statistics.
Measurements are presented both at vertical track incidence, and at an inclination of about $\alpha = \SI{19.3}{\degree}$, emulating the angle of the Lorentz drift in the unirradiated sensors caused by the magnetic field of the CMS solenoid.

\begin{figure}[tbp]
  \centering
  \includegraphics[width=.5\textwidth]{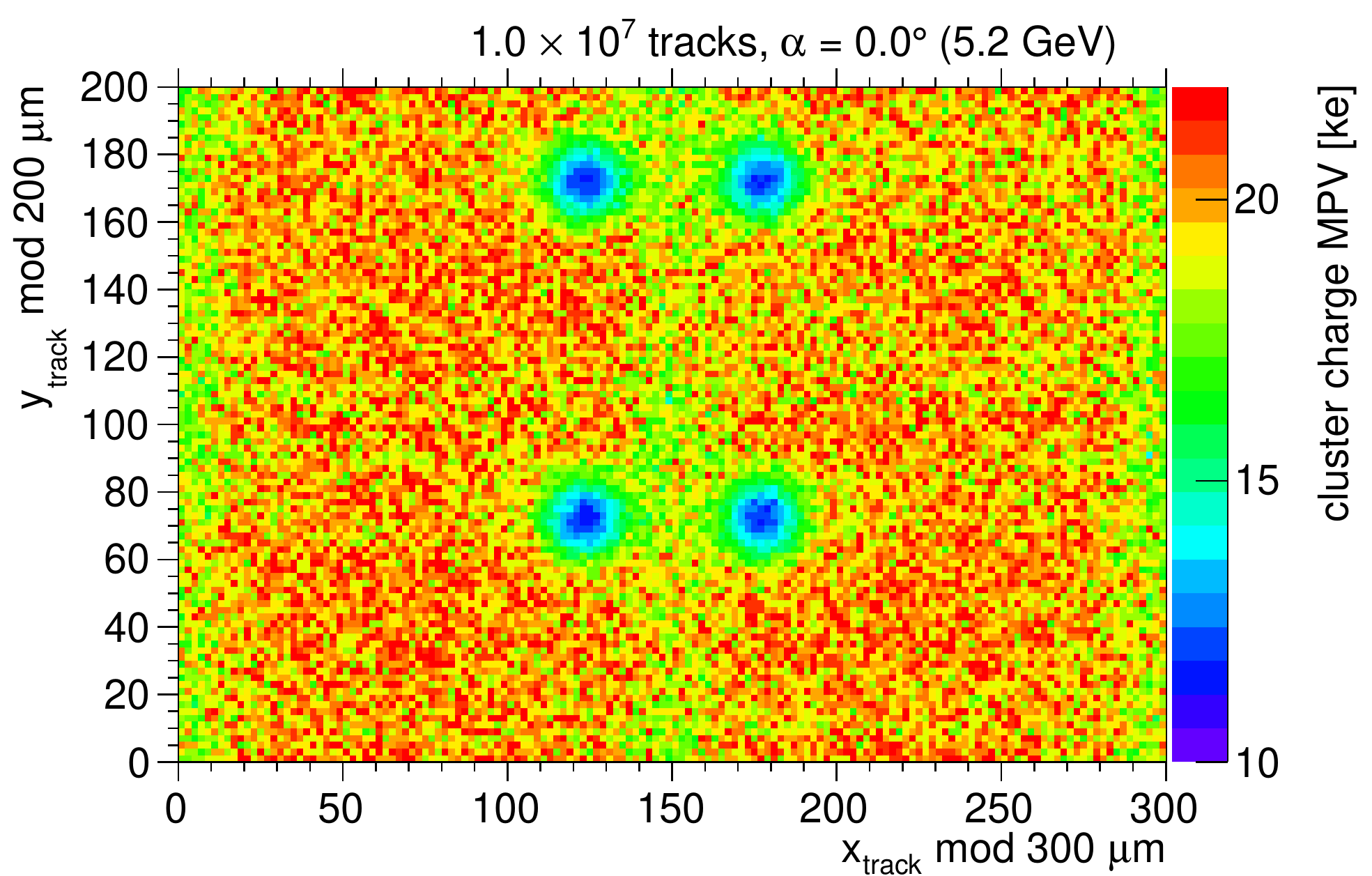}%
  \includegraphics[width=.5\textwidth]{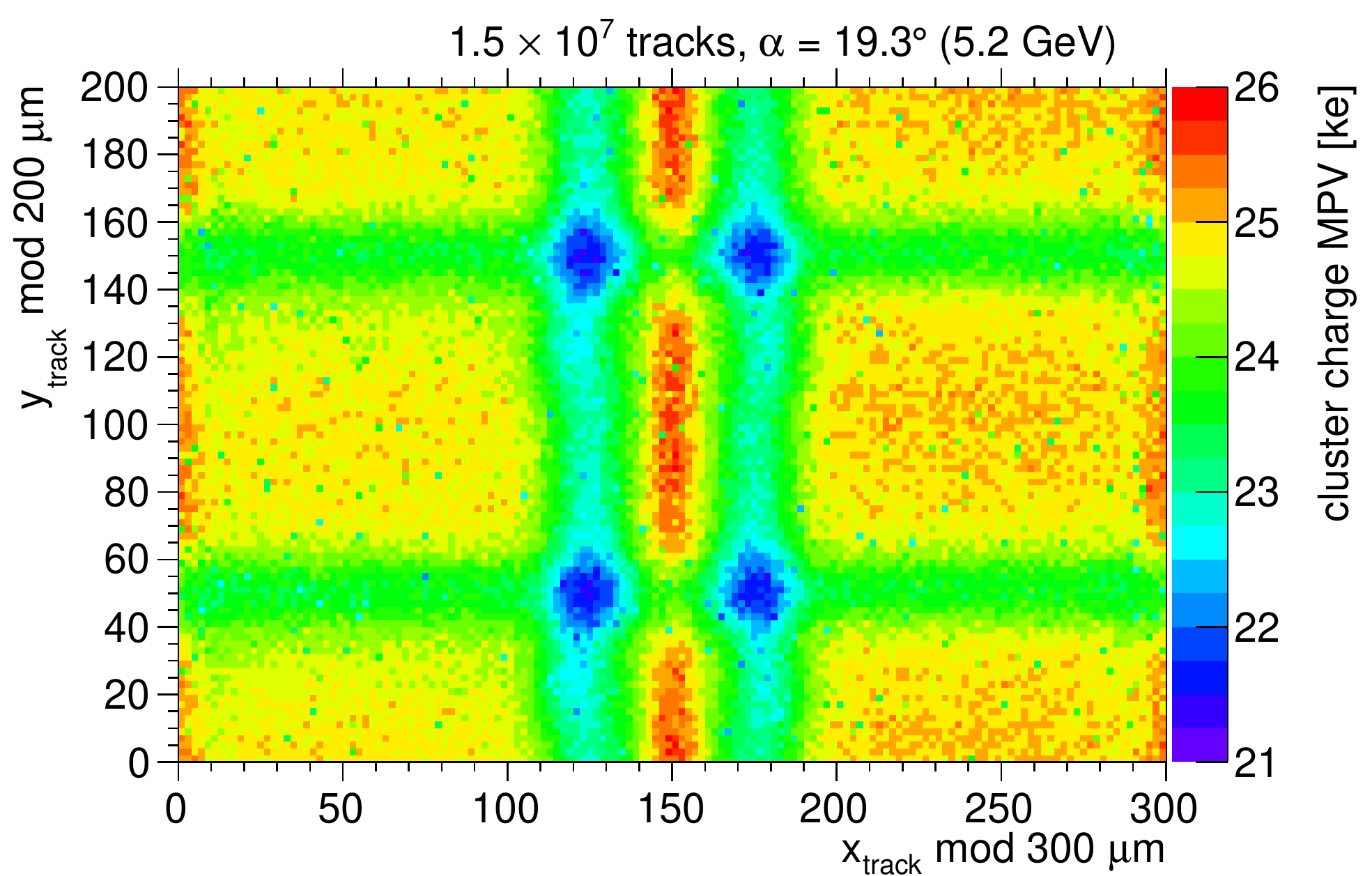}
  \caption[Charge collection in a $2\times2$ pixel array]{Charge collection in an array of $2\times2$ pixel cells. At vertical track incidence, the bias dots are clearly visible in the mean cluster charge collected, and about \SI{50}{\percent} of the total charge is locally lost~(left). At a tilt angle of $\alpha = \SI{19.3}{\degree}$, the charge loss caused by the bias dots is smeared out and reduced in magnitude, with only about \SI{15}{\percent} of the charge being lost~(right).}
  \label{fig:cce}
\end{figure}

Figure~\ref{fig:cce} shows the MPV of the cluster charge distribution as a function of the track impact position for the two different incidence angles.
A strong effect around the bias dots can be observed for the vertical track incidence caused by the altered electric field in the sensor below the bias dots on ground potential.
The cluster charge collection deficit for tracks incident vertically on the bias dot is about \SI{50}{\percent}, while the effect is smeared out at inclined incidence, where only \SI{15}{\percent} of the charge is lost.
A different scale has been chosen for the $z$ axes of the two figures in order to emphasize the features.
Neither the comparatively large charge deficit at the bias dot for vertical track incidence, nor the smeared out bias dot structure around the Lorentz angle affect the tracking performance.

\begin{figure}[tbp]
  \centering
  \includegraphics[width=.5\textwidth]{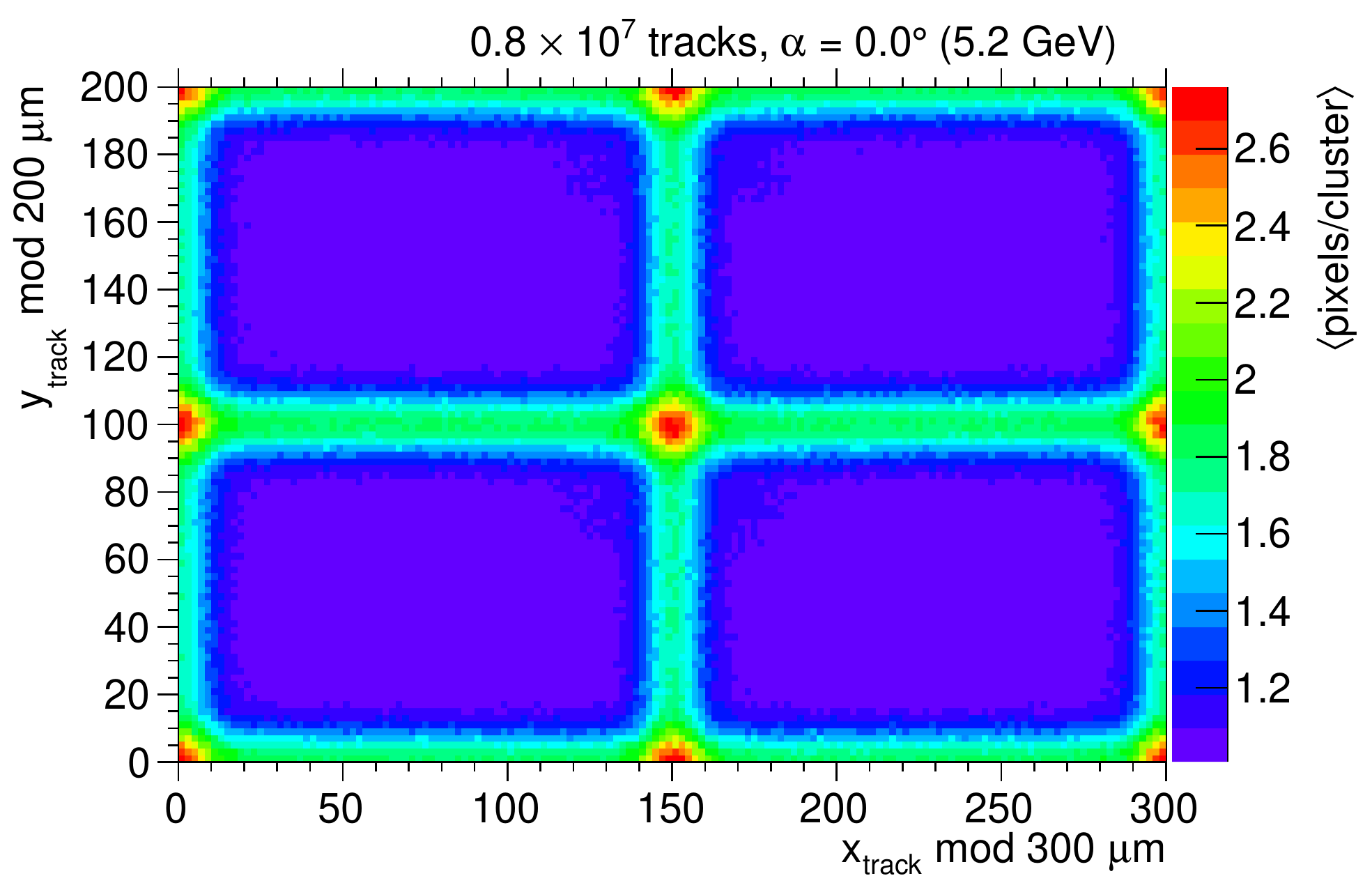}%
  \includegraphics[width=.5\textwidth]{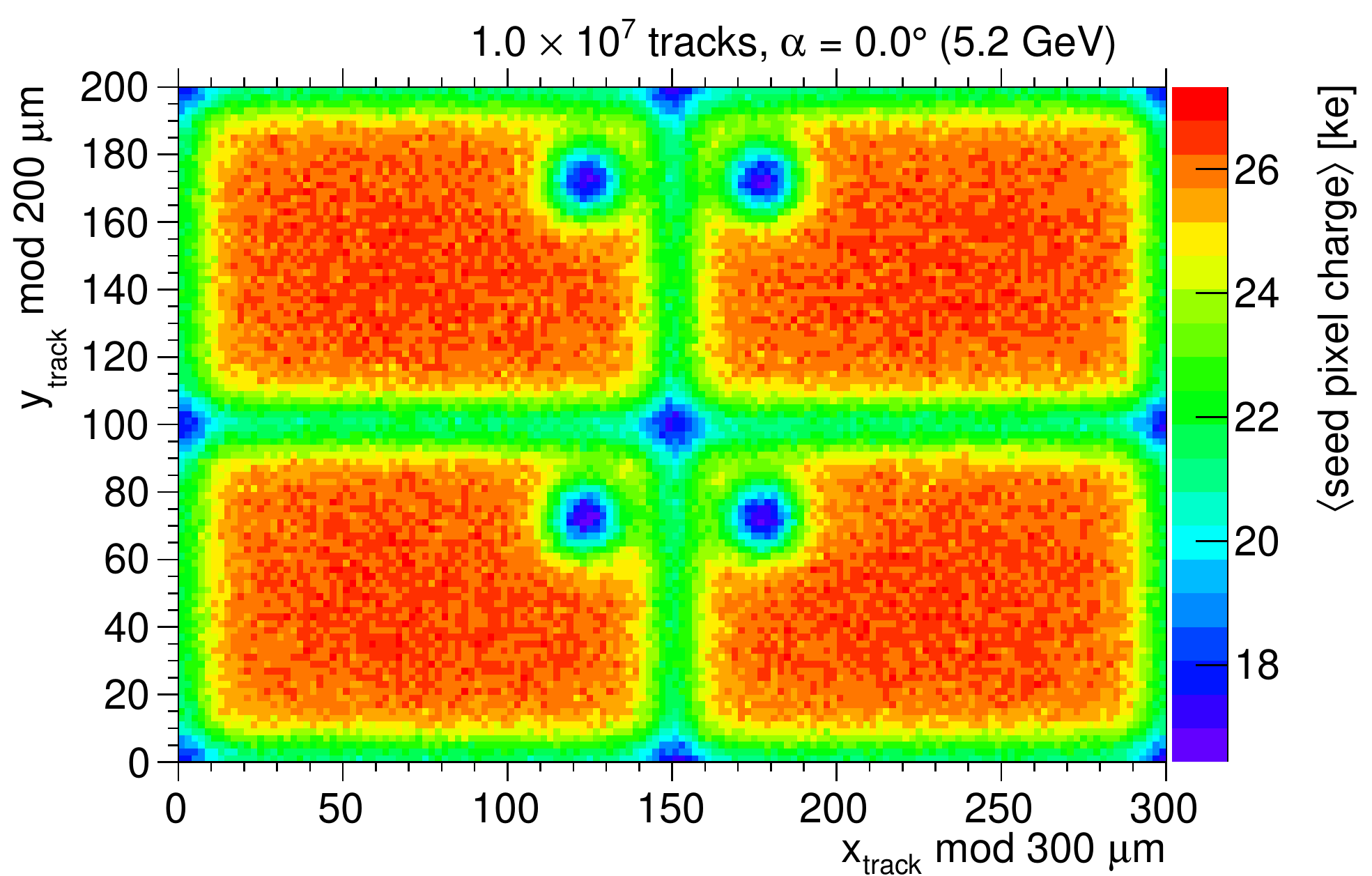}
  \caption[Mean cluster size and mean seed pixel charge in a $2\times2$ pixel array]{Mean cluster size at vertical track incidence within a $2\times2$ pixel array~(left). The regions of charge sharing are indicated by a larger mean cluster size. Mean charge of the cluster seed pixel as a function of the track impact position at vertical track incidence~(right). The implant regions can be distinguished from the \emph{p}-spray isolation and the bias dots.}
  \label{fig:cce:seed}
\end{figure}

The different regions of charge sharing between adjacent pixel cells are illustrated in Figure~\ref{fig:cce:seed}~(left) using the average cluster size as a function of the track impact position.
At vertical track incidence, the centers of the pixel cells are dominated by single-pixel clusters, while two- and three-pixel clusters mostly occur towards the edges and corners of the pixel cells, respectively.
When plotting the charge of the cluster seed pixel instead of the full cluster charge, the effect of the implant isolation structures on the charge collection can be observed as demonstrated in Figure~\ref{fig:cce:seed}~(right).
Here, the seed pixel is defined as the pixel identified using the telescope track, and its mean charge is plotted against the intra-pixel track position.
The bias dots as well as the \emph{n$^+$} pixel implants are clearly visible and are consistent with previous measurements of similar sensors~\cite{rohe-cce}.

\begin{figure}[tbp]
  \centering
  \includegraphics[width=.5\textwidth]{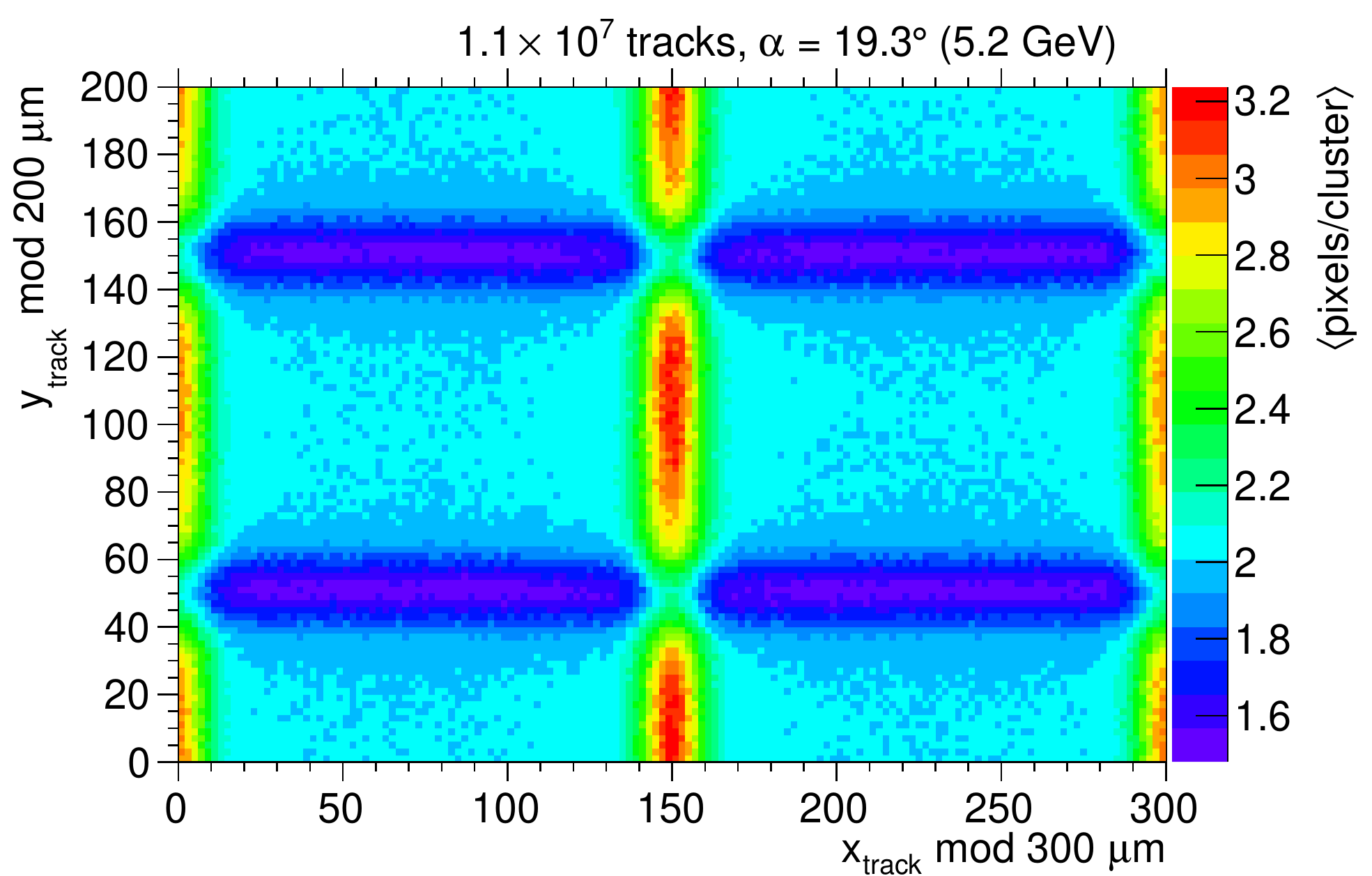}%
  \includegraphics[width=.5\textwidth]{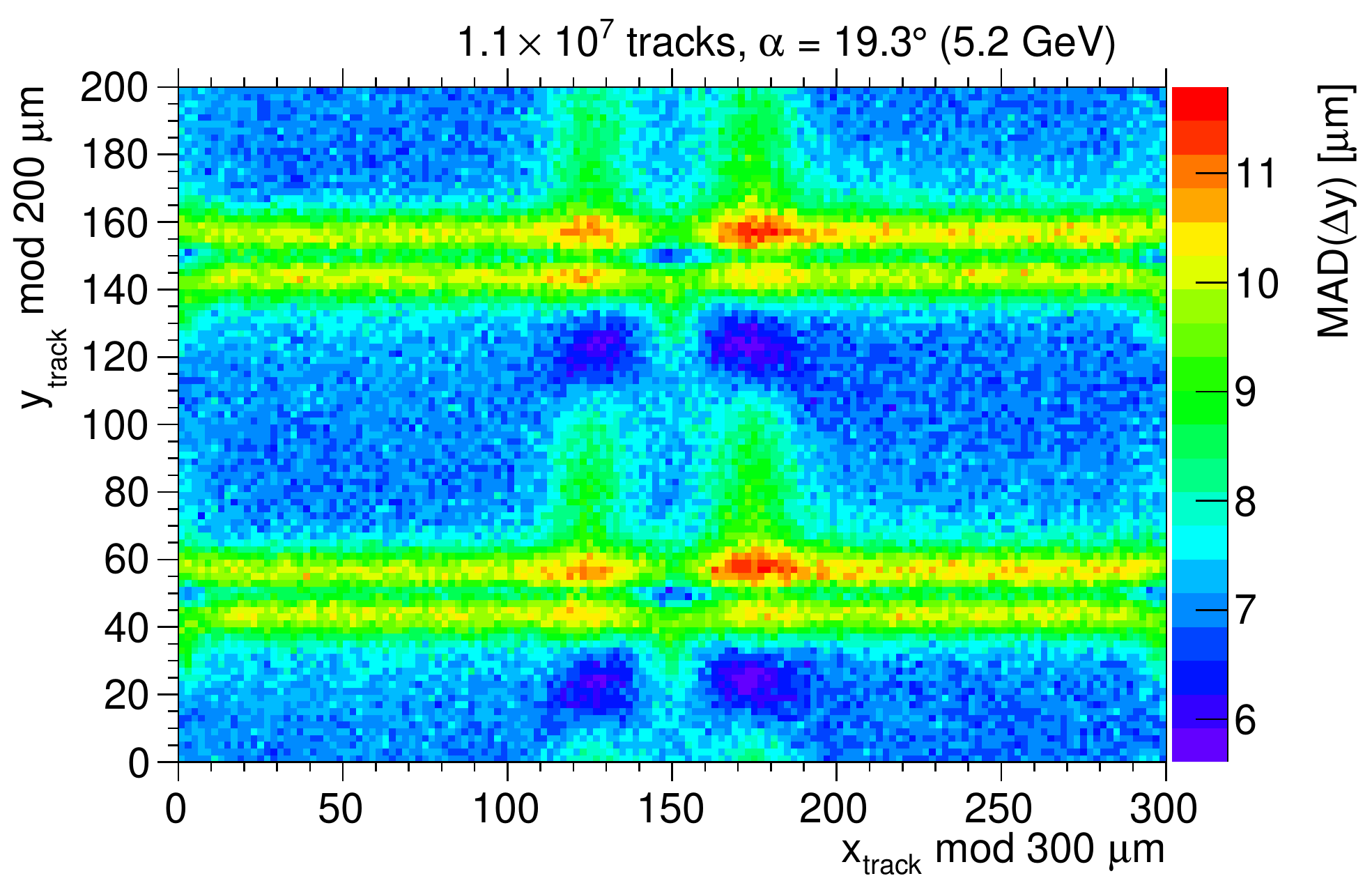}
  \caption[Mean cluster size and intra-pixel resolution in $y$ at an inclined track incidence]{Mean cluster size~(left) and intra-pixel resolution in~$y$ (residual MAD)~(right), both at an inclined track incidence of $\alpha = \SI{19.3}{\degree}$. At this angle, the bias dot deteriorates the resolution in the respective area of the pixel.}
  \label{fig:rmsy}
\end{figure}

The achieved position resolution depends on the track impact position within the pixel and the charge sharing between adjacent cells. 
With a track incidence angle of about $\alpha = \SI{19.3}{\degree}$, the cluster size distribution is dominated by two-pixel clusters as shown in Figure~\ref{fig:rmsy}~(left), with single-pixel clusters only occuring in the very center of the cells.
Figure~\ref{fig:rmsy}~(right) shows the resolution in the $y$ direction as a function of the particle track position for an incidence at $\alpha = \SI{19.3}{\degree}$.
Variations of the resolution can be observed, depending on the track impact position within the four pixels shown.
While the single-pixel clusters in the center of the pixel cells result in a good spatial resolution, threshold effects create the visible bands for particle passages that are slightly off-center.
A slight deterioration of the resolution can be observed in the region of the smeared bias dots.
Comparing to Figure~\ref{fig:cce}, the effect occurs in the regions with about \SI{15}{\percent} charge loss induced by the bias dots.
The resolutions presented are not to be confused with the actual intrinsic detector resolution described in the next section, as no track uncertainty contribution has been subtracted, and the resolution is simply calculated as the MAD of the residual distributions for convenience.


\section{Intrinsic Spatial Resolution}
\label{sec:resolution}

Comprehensive studies have been performed in order to quantify the achievable spatial resolution at the various positions in the barrel and endcap regions of the CMS pixel detector.
The intrinsic resolution is measured by fitting the residual distribution with a generalized error function, and by subtracting the telescope track pointing resolution in quadrature from the obtained distribution width.
The intrinsic resolution is measured as a function of three parameters, namely the CMS Lorentz angle (which is roughly equivalent to the tilt angle, $\alpha$, for unirradiated sensors), the position of the sensor within the barrel pixel detector
, and the pixel charge threshold.

\subsection{Resolution as a Function of the Tilt Angle}
\label{sec:lorentz}

With the \SI{3.8}{\tesla} magnetic field present in the CMS experiment, the free charge carriers produced in the silicon sensors experience the Lorentz force while drifting to the pixel implants.
For unirradiated sensors, the resulting Lorentz angle can be mimicked by tilting the sensor by the angle $\alpha$ in order to spread the charge over several rows of sensor implants.
The different path length is corrected for by normalizing the cluster charge to vertical incidence as described in Section~\ref{sec:reco}.
The pixel pitch of the sensor has been designed such that the initial Lorentz angle yields the optimal charge sharing, while increasing radiation-induced damage and underdepletion will reduce the effective charge sharing.

\begin{figure}[tbp]
  \centering
  \includegraphics[width=.5\textwidth]{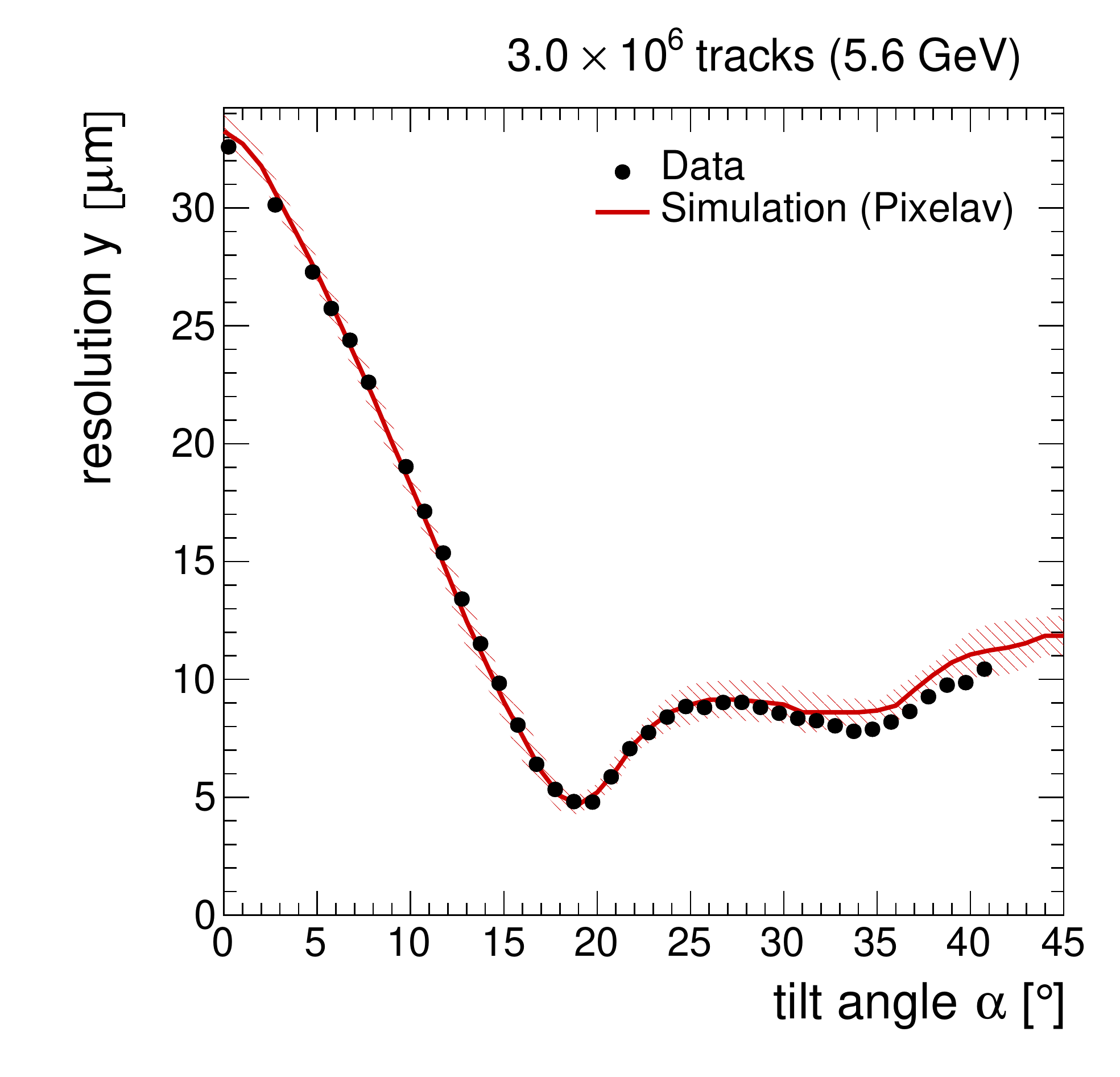}
  \caption[Position resolution as a function of the Lorentz angle]{Position resolution as a function of the tilt angle $\alpha$. In CMS, this angle denotes the $r\phi$-direction and is equivalent to the Lorentz angle induced by the magnetic field. The hatched band represents the modeling uncertainty determined from simulation.}
  \label{fig:cmsdyvstilt504}
\end{figure}

Figure~\ref{fig:cmsdyvstilt504} presents the position resolution determined using an angular scan, with the telescope track resolution subtracted quadratically.
The pronounced minimum at the optimal charge sharing angle is clearly visible, but also a local minimum at three-pixel charge sharing can be observed.
The best resolution is obtained at $\alpha = \SI{19.6}{\degree}$, where it is found to be
$$ \sigma_y = \SIERR{4.80}{0.08}{0.24}{\micro\meter}\text{,}$$
with the uncertainties calculated as detailed in Section~\ref{sec:uncertainties}.
The simulation described in Section~\ref{sec:pixelav} is shown along with the data and exhibits a good agreement.
The Lorentz angle is expected to decrease with increasing radiation-induced damage and the achieved spatial resolution is thus subject to the steep rise visible for smaller values of $\alpha$.

\subsection{Resolution as a Function of the Pseudorapidity}
\label{sec:eta}

Rotating the DUT assembly in the test beam around the column direction allows us to mimic different track dip angles $\omega$, and provides the possibility of studying the expected detector performance at different positions along the barrel of the CMS pixel detector.

\begin{figure}[tbp]
  \centering
  \includegraphics[width=.5\textwidth]{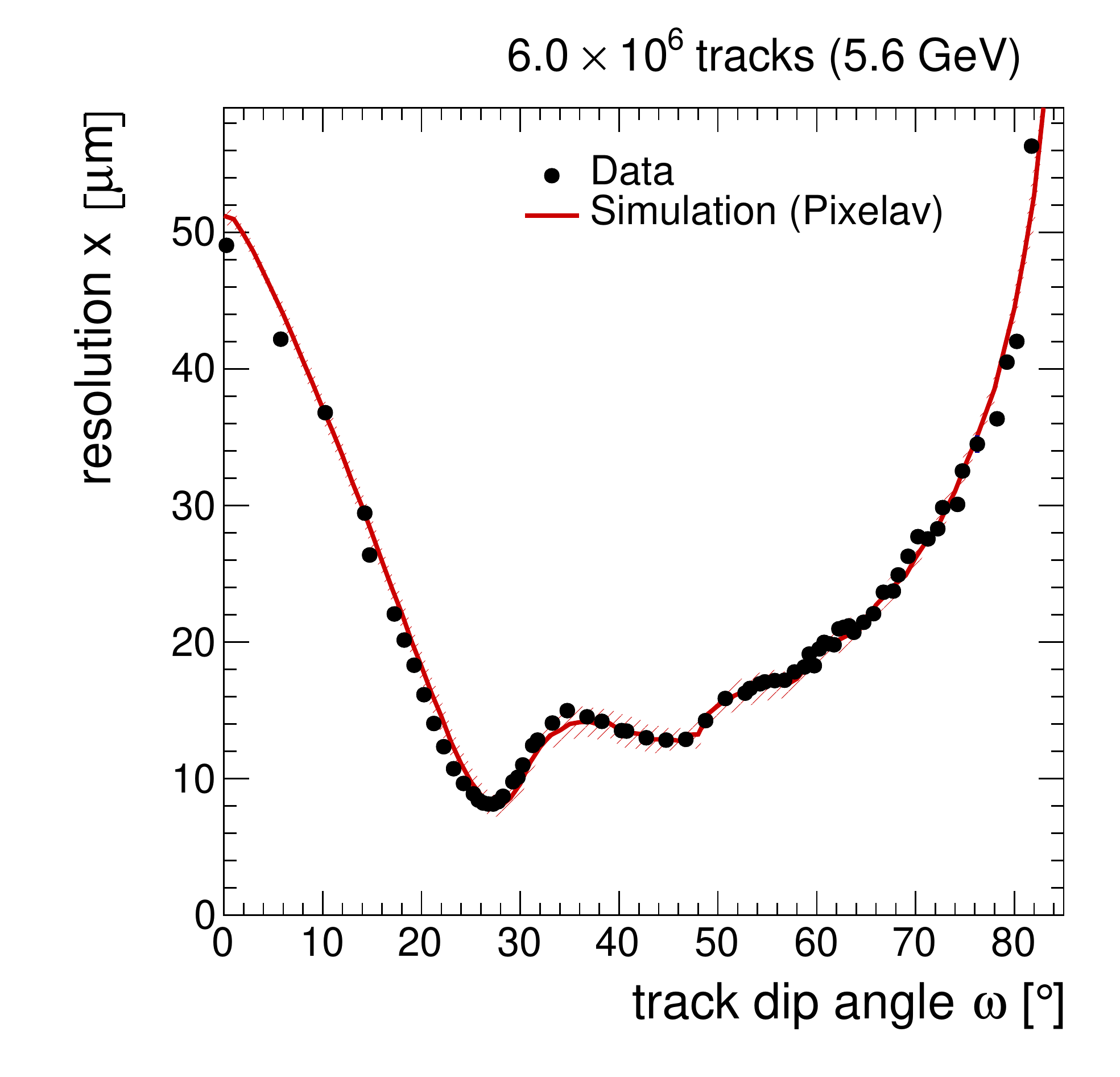}%
  \includegraphics[width=.5\textwidth]{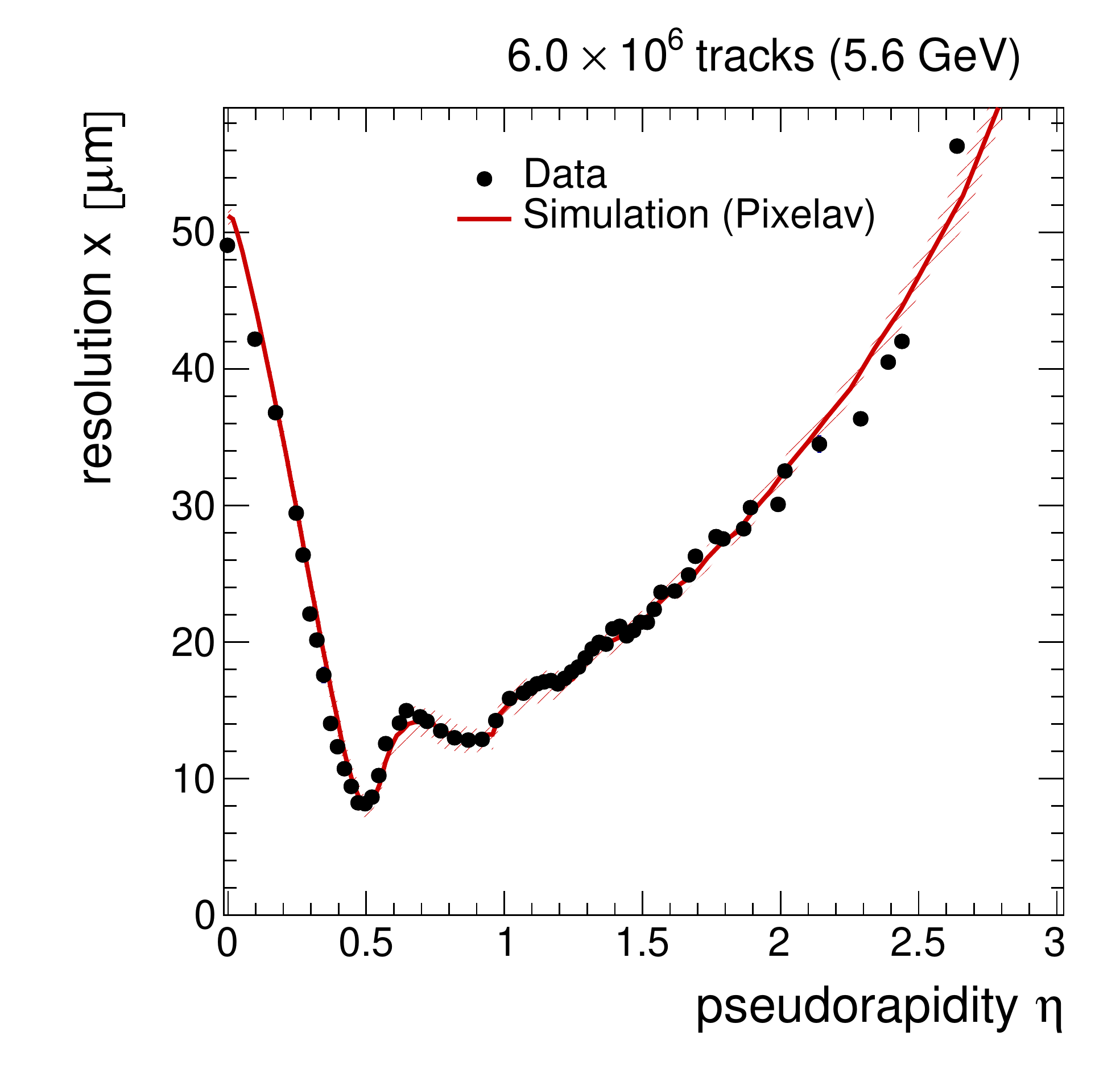}%
  \caption[Position resolution as a function of the pseudorapidity]{Position resolution as a function of the track dip angle $\omega$~(left) and the pseudorapidity $\eta$~(right). The minima for two-pixel and three-pixel charge sharing can be distinguished, while higher order minima are smeared out. The hatched band represents the modeling uncertainty determined from simulation.}
  \label{fig:cmsdyvstilt506}
\end{figure}

An angular scan up to $\omega = \SI{82}{\degree}$ is presented in Figure~\ref{fig:cmsdyvstilt506}, plotted both as a function of the tilt angle $\omega$ and as a function of the pseudorapidity $\eta$.
The pseudorapidity is defined as $\eta = - \ln{\tan{\theta/2}}$, where $\theta = \pi/2 - \omega$ is the polar angle, measured with respect to the beam axis of the CMS detector.
The optimal resolution is achieved at a track dip angle of $\omega = \SI{27.1}{\degree}$ ($|\eta| = 0.49$) and yields
$$\sigma_x = \SIERR{7.99}{0.09}{0.19}{\micro\meter}\text{.}$$
The simulation shown compares well with the data reproducing all features observed in the test beam measurements.
It is worthwhile pointing out that the blades of the forward pixel detector are tilted such that optimal charge sharing over two pixels with \SI{150}{\um} pitch is ensured.
Thus, the position resolution of the forward detector modules is expected to be around this optimal value, independent of their position in the disks.

\subsection{Resolution as a Function of the Charge Threshold}
\label{sec:threshold}

The pixel charge threshold is a crucial parameter for the position resolution.
With rising threshold, it is more likely that edge pixels within the cluster will not collect the charge required to pass the threshold.
This has an impact on the cluster size, and thus on the interpolated cluster position, and can be quantified by studying the detector behavior at different threshold settings.
These threshold measurements have been performed at a track incidence angle of $\alpha = \SI{19.5}{\degree}$, which is close to the optimum value described above.


\begin{figure}
  \centering
  \includegraphics[width=.5\textwidth]{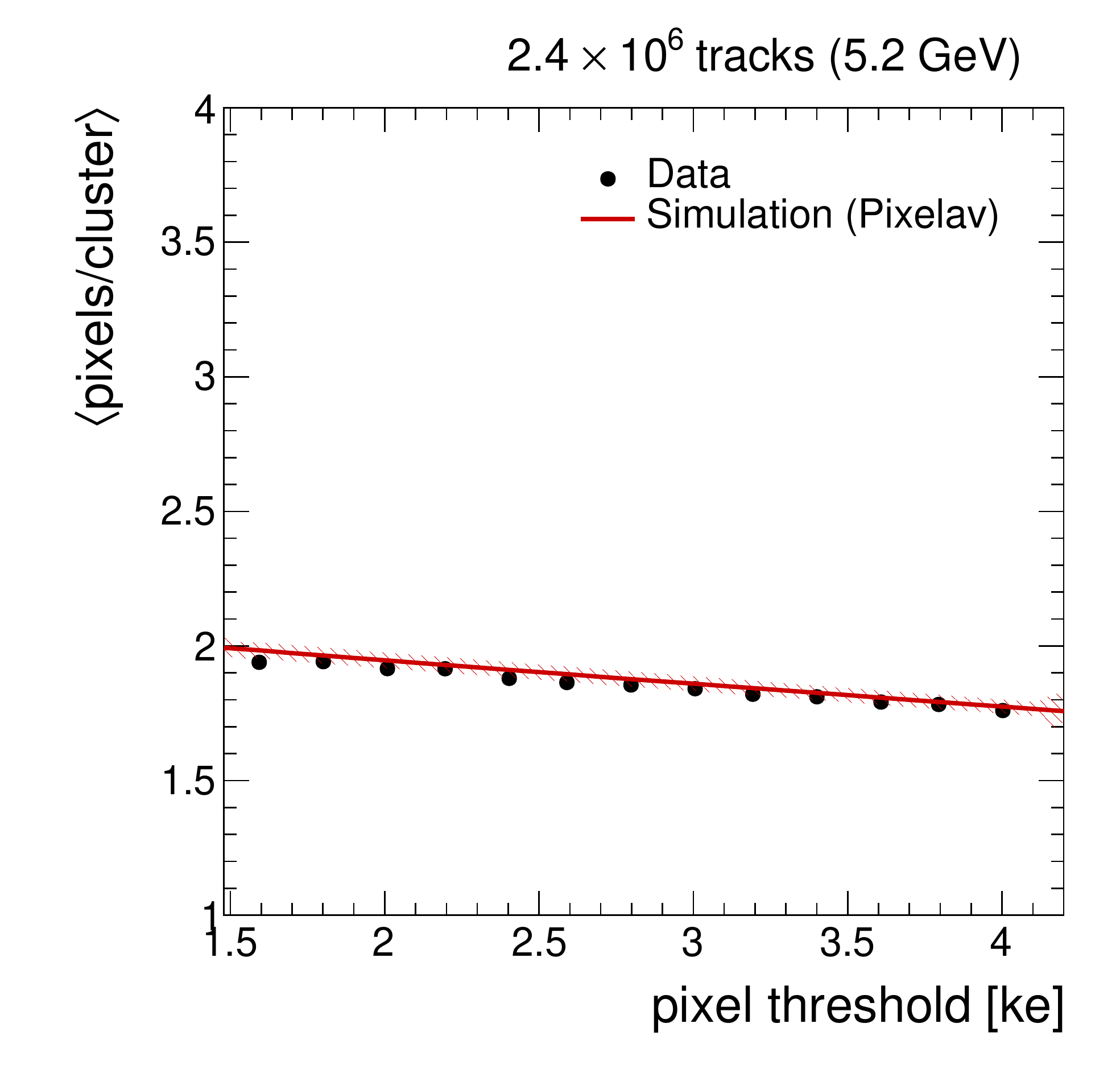}%
  \includegraphics[width=.5\textwidth]{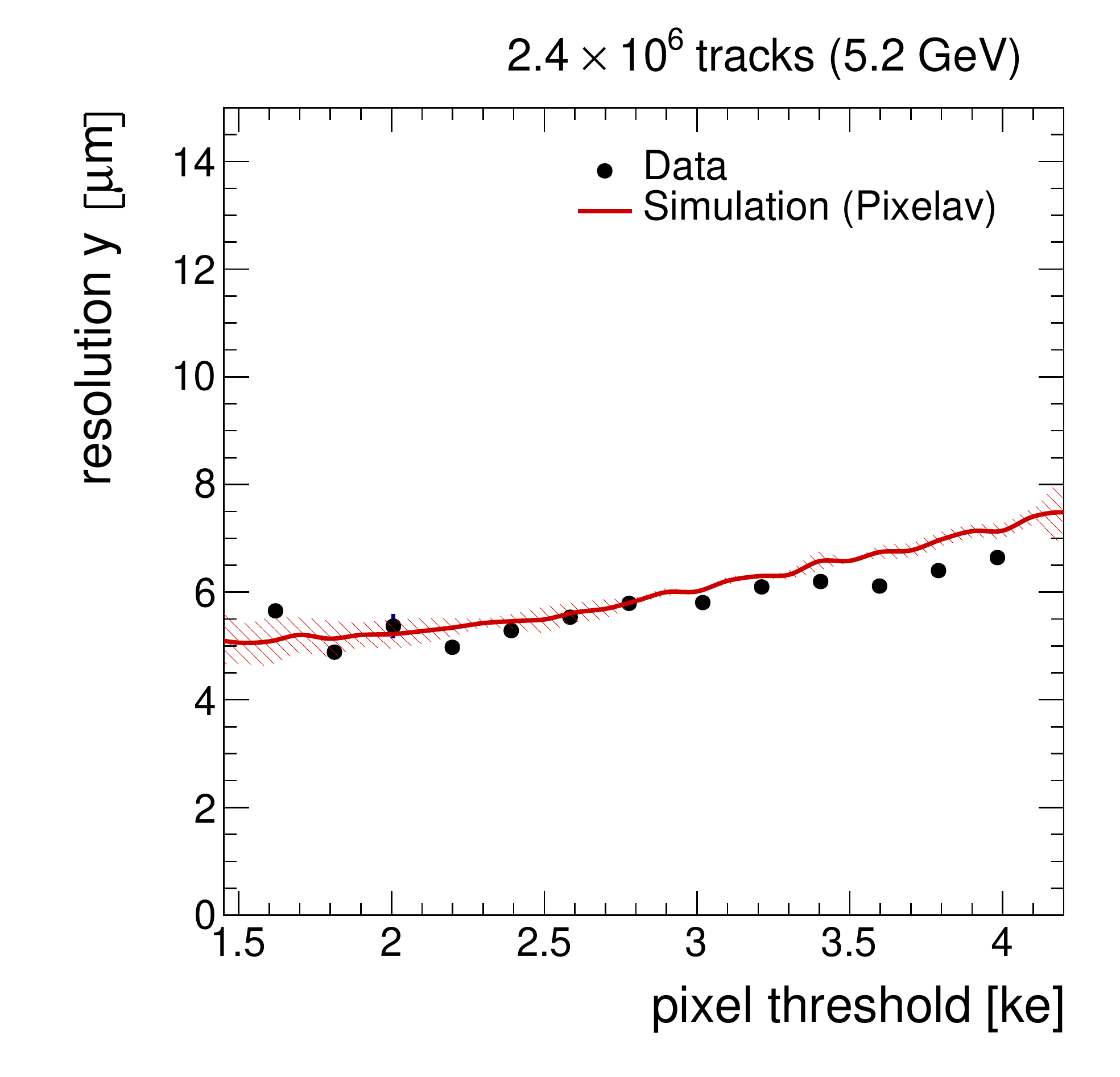}
  \caption[Cluster size and spatial resolution as function of the pixel threshold]{Mean cluster size~(left) and spatial resolution~(right) as a function of the ROC pixel threshold, for a tilt angle of $\alpha = \SI{19.5}{\degree}$. The resolution improves as the thresholds is decreased as expected. Here, the hatched band represents the modeling uncertainty associated with the thickness of the sensor in the simulation.}
  \label{fig:threshold-resolution}
\end{figure}

The effect of the charge threshold on the mean cluster size is shown in Figure~\ref{fig:threshold-resolution}~(left).
With higher thresholds the probability of losing single pixels increases, and thus the mean cluster size decreases.
The dependence of the position resolution on the charge threshold is shown in Figure~\ref{fig:threshold-resolution}~(right).
The spatial resolution is improved by about \SI{20}{\percent} by reducing the pixel charge threshold from the operation value of the ROC in the current CMS pixel detector of about \SI{3.2}{ke}~\cite{tracker-reco} to a threshold of \SI{1.7}{ke} achievable with the ROC for the Phase~I pixel detector.
The optimum resolution observed here confirms the measurement described in Section~\ref{sec:lorentz}.

\section{Conclusions}
\label{sec:conclusion}

Comprehensive test beam measurements have been performed at the DESY II synchrotron in order to characterize the behavior of the ROC intended for use with layers 2 -- 4 of the CMS Phase~I pixel detector.
The tracking efficiency is uniform over the full fiducial sensor area and has been measured to be \mbox{$\epsilon = \SIERR{99.95}{0.01}{0.05}{\percent}$} at low occupancy.

Detailed studies of various intra-pixel quantities such as resolution and tracking efficiency have been performed.
The distinct charge deficit at the bias dot at vertical track incidence is mitigated by tilting the sensor.
This scenario is close to that expected within the actual CMS experiment; the measured charge loss is observed to be only about \SI{15}{\percent}.
While the tracking efficiency is not influenced by the reduction in collected charge even at vertical incidence, small variations in the position resolution are observed around the bias dots for track incidence angles around \SI{19}{\degree}.

A strong dependence of the resolution on the cluster charge produced by delta rays has been observed.
The intrinsic spatial resolution has been measured as a function of the angles that represent rotations around columns and rows.
The minimum row resolution at the cluster charge MPV along the \SI{100}{\um} pixel pitch is $\sigma_y = \SIERR{4.80}{0.08}{0.24}{\micro\meter}$.
This corresponds to the $r\phi$ orientation within the CMS experiment where the charge sharing is determined by the Lorentz angle.
The optimal column resolution along the \SI{150}{\um} pitch side (which is equivalent to $\eta$ in CMS) is determined to be $\sigma_x = \SIERR{7.99}{0.09}{0.19}{\micro\meter}$.
Finally, the dependence of the position resolution on the pixel charge threshold has been measured.
An improvement of about \SI{20}{\percent} can be observed when reducing the threshold from \SI{3.2}{ke} to \SI{1.7}{ke} as is possible with the ROC for the Phase~I pixel detector.

Detailed simulations have been produced and compared with the test beam data.
Most of the features observed in the data are correctly reproduced by the simulation, which enables it to be used to predict future operational parameters and detector behavior.

Further studies with irradiated CMS pixel detector assemblies, investigating their performance after irradiation with protons and neutrons, are currently underway and will be published in a separate paper.
\acknowledgments

The measurements leading to these results have been performed at the Test Beam Facility at DESY Hamburg (Germany), a member of the Helmholtz Association (HGF).


\bibliographystyle{unsrt}
\bibliography{bibliography}
\end{document}

%% file: testbeam-paper-authors.tex
\affiliation[1]{Institut~f\"{u}r~Hochenergiephysik,~Wien,~Austria}
\author[1]{M.~Dragicevic}
\author[1]{M.~Friedl}
\author[1]{J.~Hrubec}
\author[1]{H.~Steininger}

\affiliation[2]{Helsinki~Institute~of~Physics,~Helsinki,~Finland}
\author[2]{A.~G\"{a}dda}
\author[2]{J.~H\"{a}rk\"{o}nen}
\author[2]{T.~Lamp\'{e}n}
\author[2]{P.~Luukka}
\author[2]{T.~Peltola}
\author[2]{E.~Tuominen}
\author[2]{E.~Tuovinen}
\author[2]{A. Winkler}

\affiliation[3]{Department~of~Physics,~University~of~Helsinki,~Helsinki,~Finland}
% ...

%% file: telescope.tex
\tikzset{%
  >=latex
}
\begin{tikzpicture}
  \node at (-6.75,1.5) {\textbf{Plane}};
  \draw [cyan, line width=0.2cm] (5, -0.5) -- (5, 0.5);
  \node at (6,.75) {Scintillators};
  \draw [cyan, line width=0.2cm] (-6, -0.5) -- (-6, 0.5);
  \draw [gray, line width=0.1cm] (-2.5, -1) -- (-2.5, 1);
  \node at (-2.5,1.5) {3};
  \draw [gray, line width=0.1cm] (-4, -1) -- (-4, 1);
  \node at (-4,1.5) {4};
  \draw [gray, line width=0.1cm] (-5.5, -1) -- (-5.5, 1);
  \node at (-5.5,1.5) {5};
  \node at (-4,-1.5) {\emph{downstream}};
  \draw [gray, line width=0.1cm] (2.5, -1) -- (2.5, 1);
  \node at (2.5,1.5) {2};
  \draw [gray, line width=0.1cm] (3.5, -1) -- (3.5, 1);
  \node at (3.5,1.5) {1};
  \draw [gray, line width=0.1cm] (4.5, -1) -- (4.5, 1);
  \node at (4.5,1.5) {0};
  \node at (3.5,-1.5) {\emph{upstream}};
  \draw [purple, line width=0.075cm, ->] (6.5,0) -- (-7.75,0);
  \node at (6.75,-.5) {Test beam};
  \draw [teal, line width=0.2cm] (-0.5, -0.75) -- (0.5, 0.75);
  \node at (0,-1.5) {\textbf{DUT}};
  \draw [black, <->] (0.25, -0.5) -- (2.25, -0.5);
  \node at (1.25,-0.75) {$dz_{\textrm{DUT}}$};
  \draw [line width=0.15cm] (-7, -0.75) -- (-7, 0.75);
  \node at (-7,-1.5) {\textbf{REF}};
\end{tikzpicture}